\documentclass[conference]{IEEEtran}
\IEEEoverridecommandlockouts

\usepackage{cite}
\usepackage{amsmath,amssymb,amsfonts}
\usepackage{graphicx}
\usepackage{textcomp}

\usepackage{url}
\usepackage{svg}
\usepackage{tikz}
\usepackage{float}
\usepackage{subfig}
\usepackage{tabulary}
\usepackage{makecell}
\usepackage[utf8]{inputenc}
\usepackage{enumitem}
\usepackage{balance}
\usepackage{color}
\usepackage{xcolor}
\usepackage{soul}
\usepackage{array}
\usepackage{caption}
\usepackage{algpseudocode}
\usepackage{algorithmicx}
\usepackage{listings} 
\usepackage{titlesec}
\usepackage{flushend} 
\newcolumntype{K}[1]{>{\centering\arraybackslash}p{#1}}
\newcolumntype{P}[1]{>{\raggedright\arraybackslash}p{#1}}
\graphicspath{{figures/}}

\def\BibTeX{{\rm B\kern-.05em{\sc i\kern-.025em b}\kern-.08em
    T\kern-.1667em\lower.7ex\hbox{E}\kern-.125emX}}
\begin{document}

\title{Optimizing Distributed ML Communication with Fused Computation-Collective Operations}

\author{\IEEEauthorblockN{Kishore Punniyamurthy}
\IEEEauthorblockA{Kishore.Punniyamurthy@amd.com}
\and
\IEEEauthorblockN{Khaled Hamidouche}
\IEEEauthorblockA{khaledhamidouche@gmail.com}
\and
\IEEEauthorblockN{Bradford M. Beckmann}
\IEEEauthorblockA{Brad.Beckmann@amd.com}
}

\maketitle

\begin{abstract}
In order to satisfy their ever increasing capacity and compute requirements, machine learning models are distributed across multiple nodes using numerous parallelism strategies. As a result, collective communications are often on the critical path, and hiding their latency by overlapping kernel-granular communication and computation is difficult due to the absence of independent computation. 

In this work, we propose fusing computation with dependent collective communication by leveraging GPUs' massive parallelism and GPU-initiated communication. We have developed self-contained GPU kernels where threadblocks/workgroups (WGs) immediately communicate their results to remote GPUs when they complete their computation. Meanwhile, other WGs within the same kernel perform overlapping computation, maintaining high ALU utilization. Such fine-grain overlapping provides the additional benefit that peak network bandwidth demand is reduced and communication is spread across the entire lifetime of application rather than only at kernel boundaries. Furthermore, we propose zero-copy optimizations for scale-up communication where the data computed by one GPU is directly communicated to peer GPUs, eliminating intermediate stores and buffering.

We demonstrate our approach by creating three prototype fused operators (embedding + All-to-All, GEMV + AllReduce, and GEMM + All-to-All) to address the pervasive communication overheads observed in deep learning recommendation models (DLRM), Transformers and Mixture of Experts  (MoE) model architectures.
In order to demonstrate that our approach can be integrated into ML frameworks for wide adoption in production environments, we expose our fused operators as new PyTorch operators as well as extend the Triton framework to enable them.
Our evaluations show that our approach can effectively overlap communication with computations, subsequently reducing their combined execution time than the current collective library-based approaches. Our scale-up GEMV + AllReduce and GEMM + All-to-All implementations achieve 13\% (up to 22\%) and 12\% (up to 20\%) lower execution time, while our fused embedding + All-to-All reduces execution time by 20\% and 31\% for intra-node and inter-node configurations. Large scale-out simulations indicate that our approach reduces DLRM execution time by 21\% for 128 node system.
\end{abstract}


\section{Introduction}
Machine learning is currently being used across a wide variety of applications ranging from classification (e.g., fraud detection, medical diagnostics, facial recognition), pattern analysis (e.g., product recommendations, stock price prediction) to content generation (e.g., code generation, chatbots, image/video generation). Machine learning (ML) models are increasing in size to tackle more complex problems. Studies~\cite{mlmodel_trend} have shown that the model sizes have increased by five orders of magnitude between 2018 to 2022. Large ML models have consequently fueled the development of distributed systems capable of meeting their capacity and compute requirements. Subsequently, parallelization techniques have been developed to efficiently map ML models on to distributed systems~\cite{parallelism_strategy}. The resulting communication in distributed ML models (e.g., weight updates, activation exchanges between layers, etc.) are difficult to hide due to absence of independent computation and thereby become a significant bottleneck~\cite{meta_dlrm_neo, tutel} as the system scale grows. 


\begin{figure}[t]
\captionsetup{justification=centering}
  \centering
    \subfloat[Traditional node design]{\includegraphics[width =0.32\columnwidth]{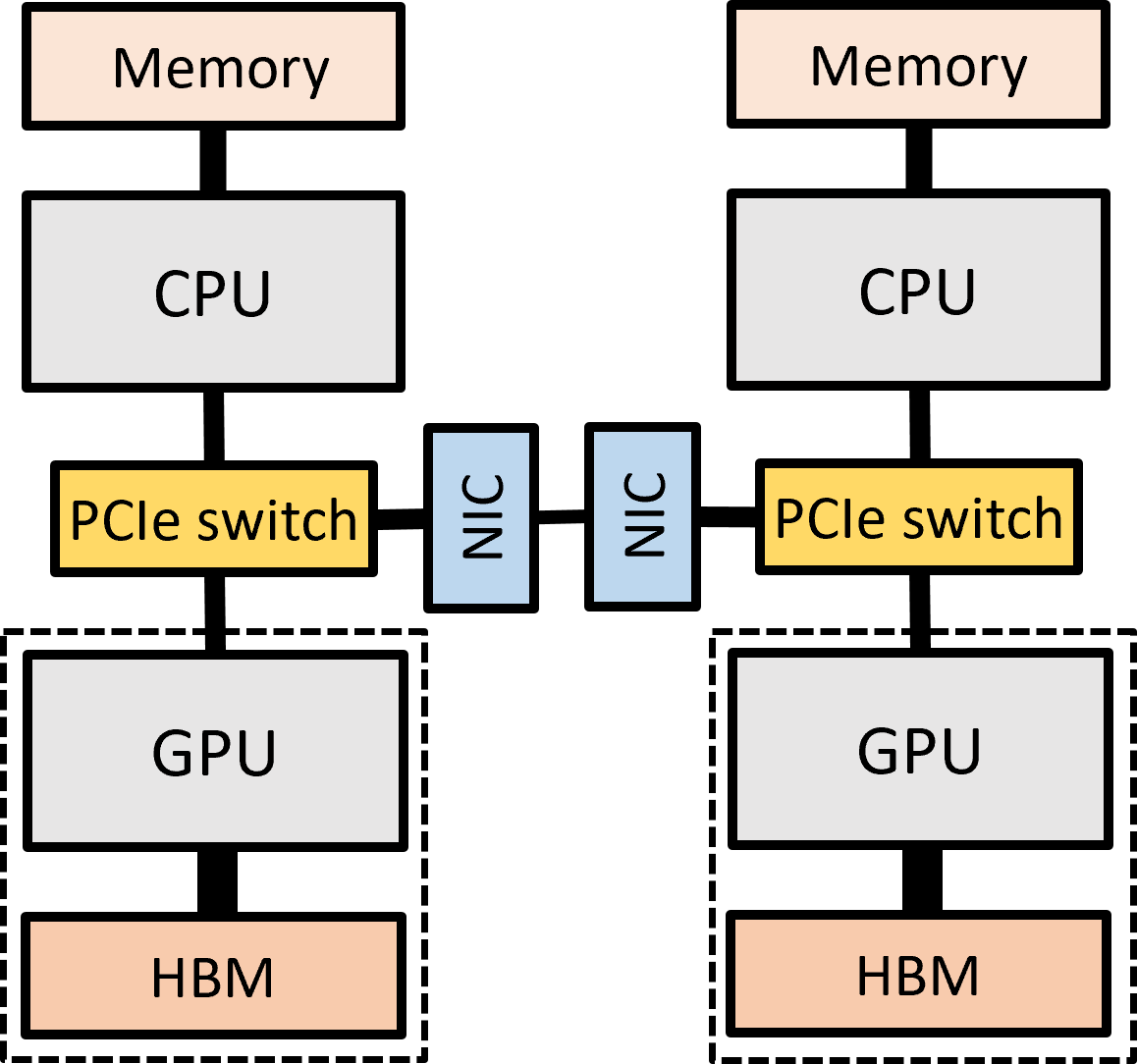}\label{fig:typical_system}}
  \hspace{2pt}
  \subfloat[Emerging HPC/ML \\node design]{\includegraphics[width =0.65\columnwidth]{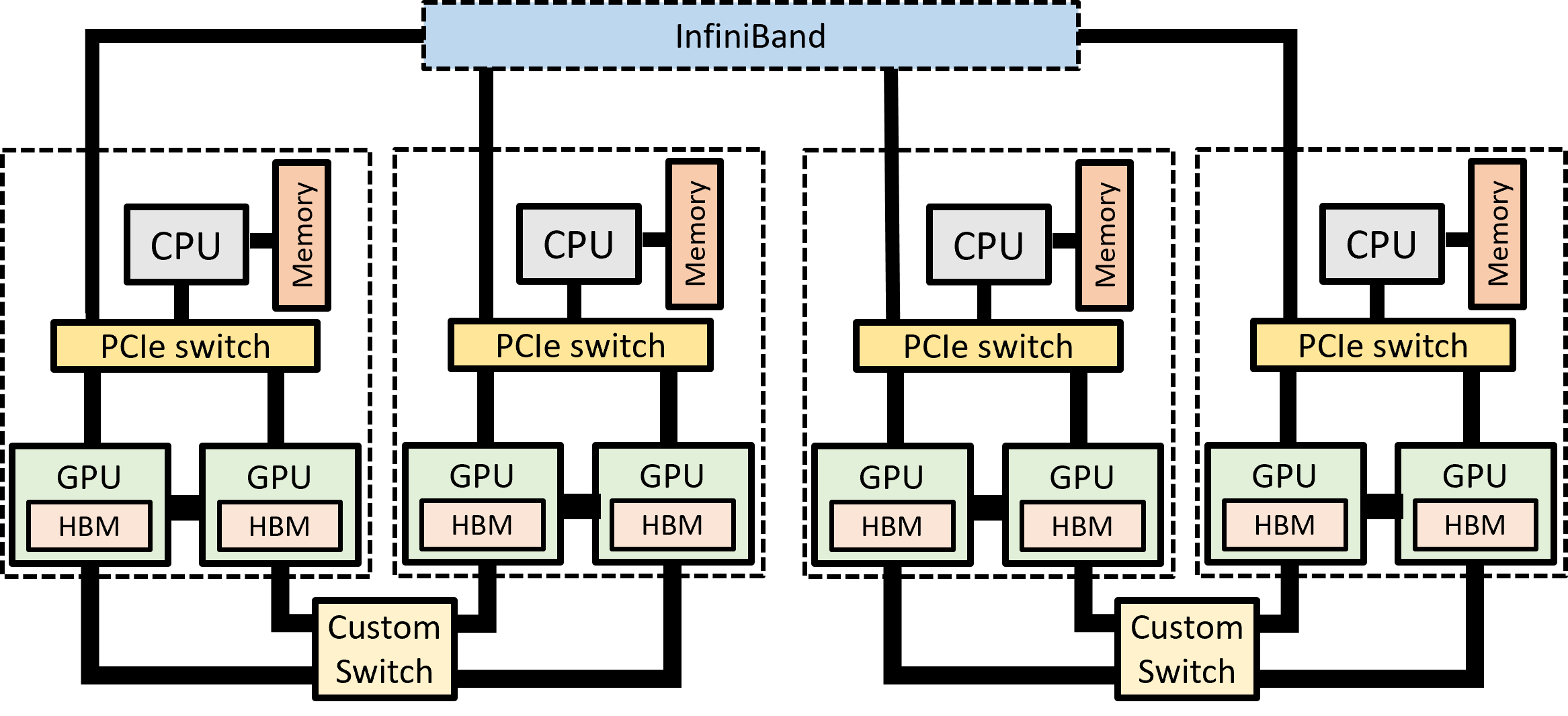}\label{fig:gh_system}}
    \caption{System architecture trends.}
    \label{fig:sys_arch_trends}
    \vspace{-15pt}
\end{figure}

ML trends are also influencing system design and architecture development. Figure~\ref{fig:sys_arch_trends} compares a traditional CPU-GPU multi-node system with a state-of-the-art node design optimized for HPC/ML. Other than compute and memory improvements, we notice two broad trends. Firstly, intra- and inter-node communication bandwidth and latency have been improved using proprietary interconnects\cite{nvlink, infinity_fabric} and high-bandwidth network interface cards (NICs). Secondly, given the increased preference to use GPUs for ML workloads, GPUs are becoming the primary computation engine tightly integrated with on-node CPUs, GPUs and 
are directly communicating with the network using high-radix switches\cite{grace_hopper_arch} or direct-attached NICs\cite{cdna_whitepaper,frontier_olcf}. However, applications have yet to fully embrace these trends and often rely on the CPU~\cite{rdma_hpc, openshmem_kvstore} to perform remote communication. While progress has been made to remove CPUs from the data path, for example, GPUDirect RDMA~\cite{gpudirect, cuda-aware_MPI} enables data to be directly moved between GPU memory and NIC, communication is still triggered by the host CPU and often at kernel boundaries. Such approaches are mostly suitable for bulk-synchronous applications where communication can be overlapped with independent computation at the kernel granularity and the kernel-launch overhead is amortized using large kernels. 

In this paper, we aim to address the growing bottleneck of collective communication by fusing and overlapping communication and dependent computation using intra-kernel GPU-initiated networking. Recent GPU micro-architectural features (e.g., GPU cache flush instructions{~\cite{MI200_isa}}, threadblock cluster{~\cite{hopper_arch}}) have made GPU-initiated networking operations performant and thus a pragmatic option. 
Our approach is easy to apply and leverages the existing threadblock/workgroup\footnote{Workgroup is the OpenCL equivalent of CUDA threadblock}-level work partitioning in GPU applications and does not require any hardware changes. We have developed self-contained, persistent GPU kernels~\cite{cta_sched_persistentkernel} where one or more logical WGs upon completing their share of computations, issue non-blocking transactions to communicate their results to remote GPUs. By immediately scheduling non-blocking network transactions, data fragments are communicated as they are computed without waiting for kernel completion. 

We demonstrate our approach by creating fused communication-computation kernels, to address the collective bottlenecks in popular ML architectures, such as deep learning recommendation models (DLRM), Transformers, and mixture of experts (MoE). Specifically, we create prototype kernels which fuse embedding pooling, matrix-vector multiplication (GEMV), matrix-matrix multiplications (GEMM) with collectives (All-to-All, and AllReduce). Our approach can be used to fuse both inter-node (e.g., RDMA) and intra-node GPU communication. For the locally communicated data, we develop zero-copy fused kernels where the GPU threads directly store their computed results to collective destination buffers. We also demonstrate the different ways such fused operators can be integrated within ML frameworks. Specifically we expose the new fused communication-computation kernel as a new operator within PyTorch{~\cite{Pytorch}} and extend \emph{Triton} (Python-like language framework to develop highly efficient GPU code{~\cite{triton}} integrated with PyTorch{~\cite{triton_pytorch}}) to include the necessary communication primitives to develop custom fused kernels. 

This paper makes the following key contributions:
\begin{itemize}
  \item  We propose a novel approach to fuse computation and collective communication within the same kernel, thus overlapping collective communication, and reducing peak bandwidth. Further, we propose zero-copy fused kernels where the results are directly written to the peer GPU memory thereby eliminating intermediate buffering and copy operations.

  \item We develop three first of their kind fused communication-computation prototype kernels to address the collective overhead in DLRM, Transformers, and MoE architectures. Specifically, we develop \emph{embedding pooling + All-to-All}, \emph{GEMV + AllReduce}, and \emph{GEMM + All-to-All} fused operators. Our evaluations show that our fused operators achieve 31\%, 13\% and 12\% lower latency respectively.

  \item We implement two different approaches to integrate fused operators within ML frameworks. 1) We expose fused communication-computation kernel as a new operator within PyTorch to be transparently used by developers. 2) We extend Triton framework to include communication primitives allowing users to develop custom fused kernels catering to their needs. In this work, we use our Triton extensions to develop \emph{GEMM + All-to-All} kernel.     

  
  \item We evaluate the trade-offs for using GPU-initiated communication and highlight the factors critical (e.g., occupancy) to achieve effective overlap with low overhead.
\end{itemize}

\begin{figure}[b]
  \vspace{-10pt}
  \centering
     \includegraphics[width =0.75\columnwidth,keepaspectratio]{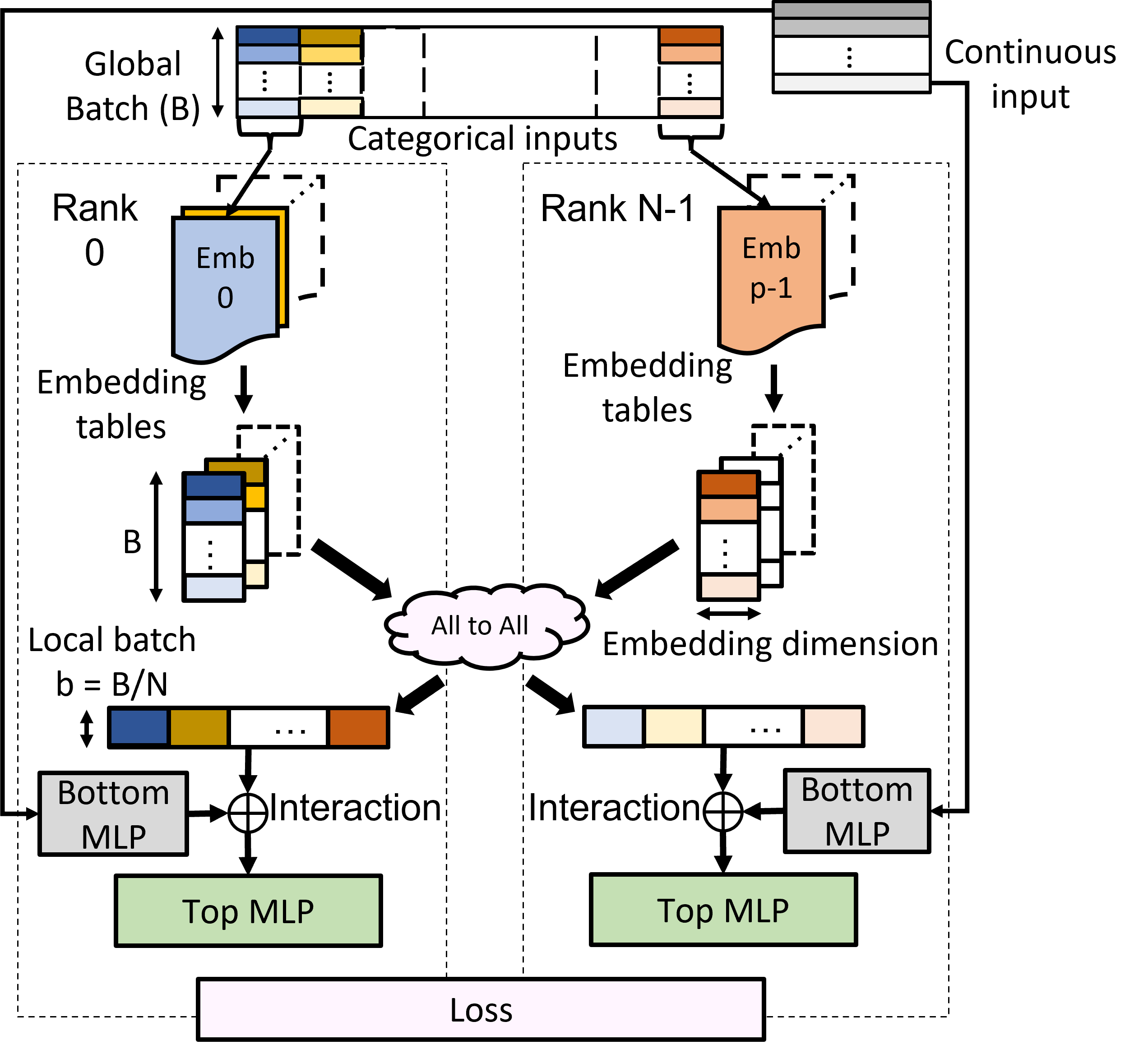}
     
    \caption{DLRM forward pass.}
    \vspace{-5pt}
    \label{fig:dlrm}
\end{figure}

\section{Background}
In this section, we first illustrate the communication bottleneck observed in several distributed ML applications: DLRM, Transformers, and MoEs. We then provide a background on GPU-initiated communication and intra-kernel networking. 

\subsection{Collectives in Distributed ML Models}
Large ML models use parallelism strategies (e.g., tensor parallelism, model parallelism, fully-sharded data parallelism (FSDP)~\cite{fsdp_strategy}) to avoid data duplication across distributed nodes. However, such strategies result in additional collective communication to train and execute the models. 

\textbf{DLRM (Embedding + All-to-All):} Recommendation models are widely deployed for ranking and click through rate predictions tasks. Due to the large memory requirements of embedding tables, DLRMs typically employ model parallelism for distributing embedding tables (table, row, and column parallelism~\cite{meta_dlrm_neo}) across GPUs~\cite{DLRM}. The top multi-layer perceptron (MLP) layers of DLRM use data parallelism for scaling across multiple GPUs. In order to switch from model parallelism execution (for embedding operations) to data parallelism (for top MLP layers), All-to-All collective operations are used~\cite{DLRM} as shown in Figure~\ref{fig:dlrm}. 
Note that bottom MLP layers are the only independent computation available to be overlapped with the All-to-All collective but are usually small and thus cannot hide communication effectively. Prior research has shown that All-to-All latency is significantly exposed ($>$35\%) and has direct impact on overall latency in state-of-the-art systems~\cite{meta_dlrm_neo}.

Hiding the All-to-All collective will require overlapping it with dependent computations (embedding pooling or top MLP), which is not possible with the existing bulk-synchronous kernel execution. In this work, we overlap All-to-All communication with the embedding pooling operator by communicating fragments of pooling output data as they are computed. Since embedding pooling operation is parallelized across independently executing WGs, output computed by single or cluster of WGs can be communicated independent of others without violating data-dependencies.

\begin{figure}[t]
  \centering
     \includegraphics[width =0.6\columnwidth,keepaspectratio]{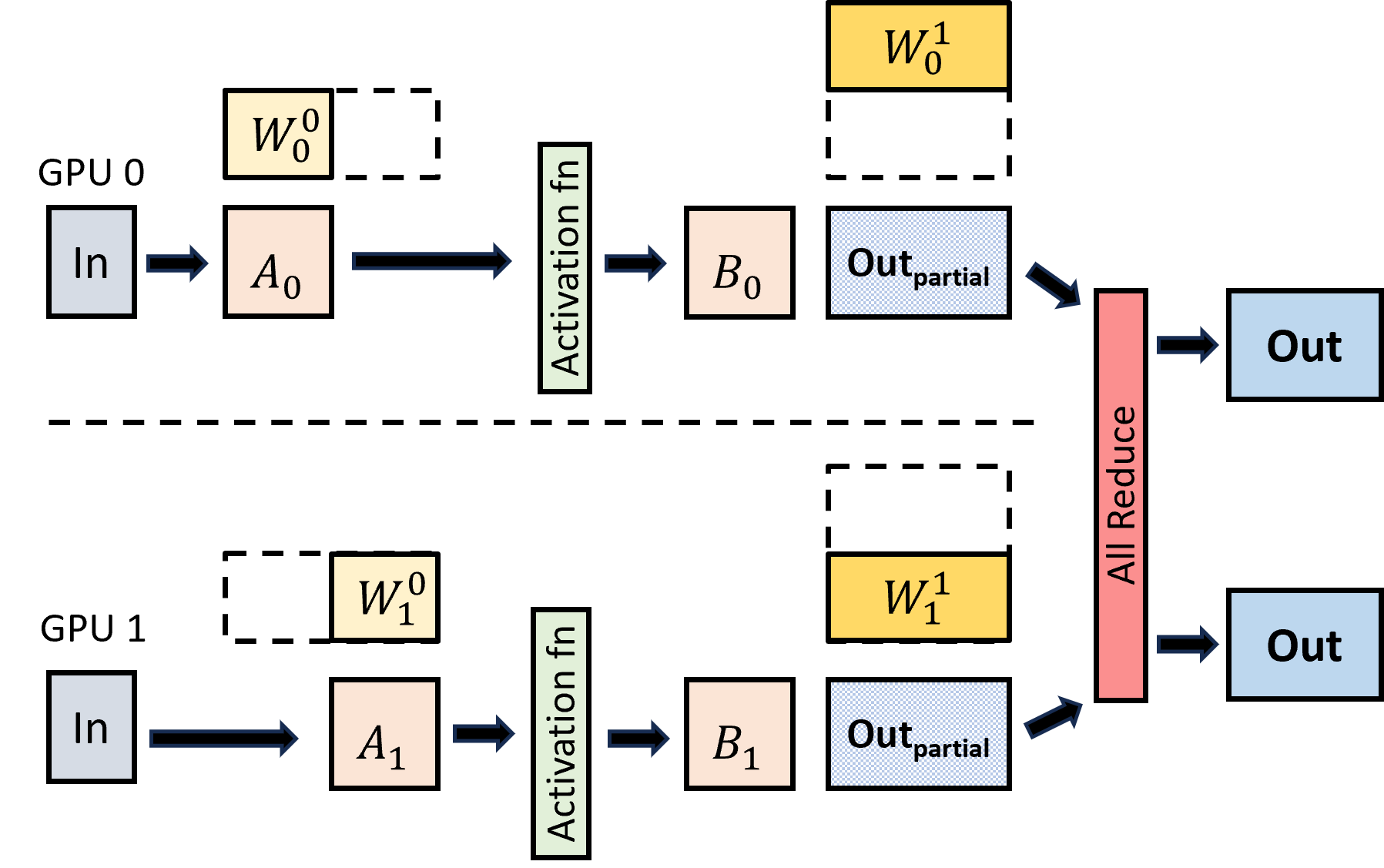}
    \caption{Model parallelism in Transformer MLP layer~\cite{megatronlm}.}
    \vspace{-10pt}
    \label{fig:megatron_gemm_ar}
\end{figure}

\textbf{Transformers (GEMV + AllReduce):} Transformers have become a popular ML model architecture for natural language processing, machine translation, and text generation tasks. Auto-regressive Transformers (e.g., GPT, Llama2) leverage the prior tokens and predict the next tokens iteratively. Their inference consists of two phases{~\cite{splitwise}} 1) \emph{Prompt (or pre-fill) phase}: all input tokens are processed to generate the first token. 2) \emph{Token (or decode) phase}: iterative phase where tokens generated previously along with the inputs are used to predict the next token. Typical models consist of a self-attention layer followed by feed-forward layer and this building block is repeatedly instantiated to create the entire model. The computations associated with these layers are matrix-matrix multiplications (during training, prompt phase of inference) or matrix-vector multiplications (token phase of inference){~\cite{T3}}, and are usually combined with element-wise operations (e.g., normalization, addition).

Transformer architectures used in modern language models are too large to fit within a single GPU device and are thus partitioned. 
Figure{~\ref{fig:megatron_gemm_ar}} shows an example of model parallelism on Transformer feed forward layers as proposed in Megatron-LM{~\cite{megatronlm}}. The feed forward layer consists of two linear layers with an activation layer in-between. The weights corresponding to each of the MLP layer are partitioned across the two GPUs, such that weights of the first layer are partitioned column-wise ($W^{0}_{0}$, $W^{0}_{1}$), while weights of second layer are partitioned row-wise ($W^{1}_{0}$, $W^{1}_{1}$). The input is duplicated across both GPUs and multiplied with half of the weights ($W^0$) in each GPU. The outputs ($A_0$, $A_1$) are passed through the activation function before being multiplied with weights of second layer ($W^1$) to generate the partial outputs ($Out_{partial}$). Finally, the partial outputs are reduced using AllReduce to obtain the final results. This AllReduce operation contributes significantly (up to 46\%{~\cite{T3}}) towards the inference latency.  

Due to lack of independent computation, the AllReduce collective can only be overlapped with the preceding computation. Since inference latency is very dominated by the token-phase{~\cite{splitwise}} which performs matrix-vector multiplication (assuming no batching), we focus on fusing AllReduce with GEMV computation in this work. GPU kernels for GEMV operation{~\cite{tensile_git}} typically divide the computation across WGs, where each WG is responsible for computing a tile of the output vector. These output tiles can be communicated and reduced independently of other tiles. We exploit this opportunity to develop a fused \emph{GEMV + AllReduce} operator.

\begin{figure}[b]
  \vspace{-10pt}
  \centering
     \includegraphics[width =0.72\columnwidth,keepaspectratio]{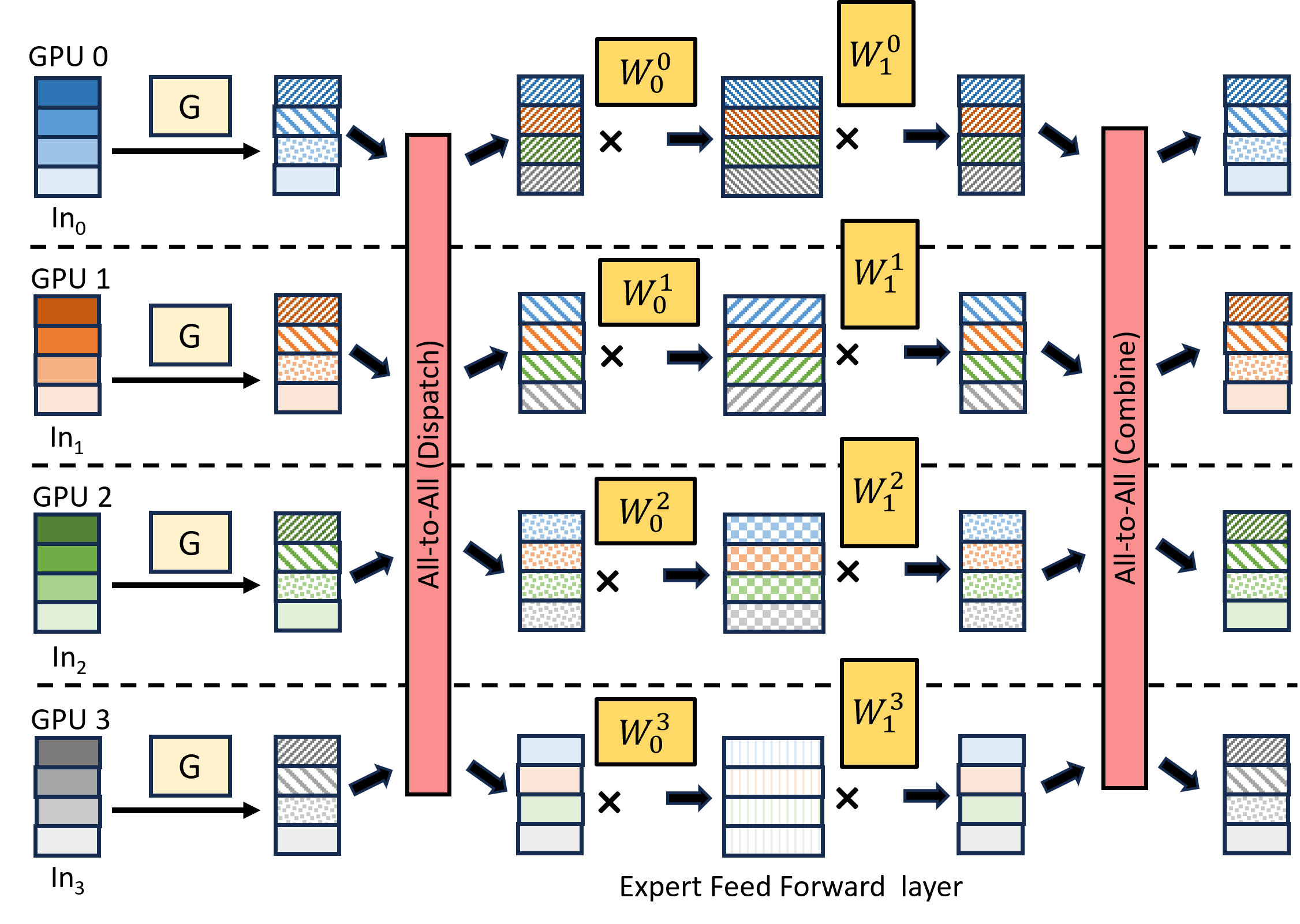}
    \caption{MoE layers distributed across GPUs~\cite{tutel}.}
    \label{fig:tutel_gemm_a2a}
\end{figure}

\textbf{Mixture of Experts (GEMM + All-to-All):} Mixture of Expert (MoE){~\cite{moe}} model architectures are growing in popularity as they allow increasing the model parameters without linearly scaling the computation cost. MoE architectures introduce parallel sub-layers called \emph{experts}, where different MoE layer inputs are passed to only a subset of experts as determined by a gating function. The expert sub-layers are typically feed-forward layers which translate to matrix-matrix multiplication (GEMM) kernels. MoEs' expert-level parallelism{~\cite{deepspeedmoe}}{~\cite{switch_transformer}} map different sub-layers across different GPUs, resulting in two All-to-All collectives, one for distributing the inputs across experts (All-to-All Dispatch) and another to gather the outputs from different experts (All-to-All Combine) as shown in Figure{~\ref{fig:tutel_gemm_a2a}}. Inputs from four GPUs ($In_{0-3}$) are distributed to the individual experts (assuming uniform distribution) based on the gating function (\emph{G}) using All-to-All (dispatch). The All-to-All output is passed as input to individual expert feed forward layer (one per GPU), the output of which is again returned to the original GPU using All-to-All (combine). These All-to-All collectives are in the critical path and adds significant overhead (up to 43\%{~\cite{tutel}}) to the MoE execution time.

The All-to-All collectives can only be overlapped with dependent GEMM operations. Similar to GEMV, WGs of the MoE GEMM kernels compute output tiles independently and thus can be communicated as soon as they are computed. In this work, we consider top-2 routing i.e. each token is communicated to two experts and assume equal workload across experts. We demonstrate our approach by fusing MoE GEMM computation with All-to-All Combine communication.

\begin{figure}[t]
  \centering
     \includegraphics[width =0.9\columnwidth,keepaspectratio]{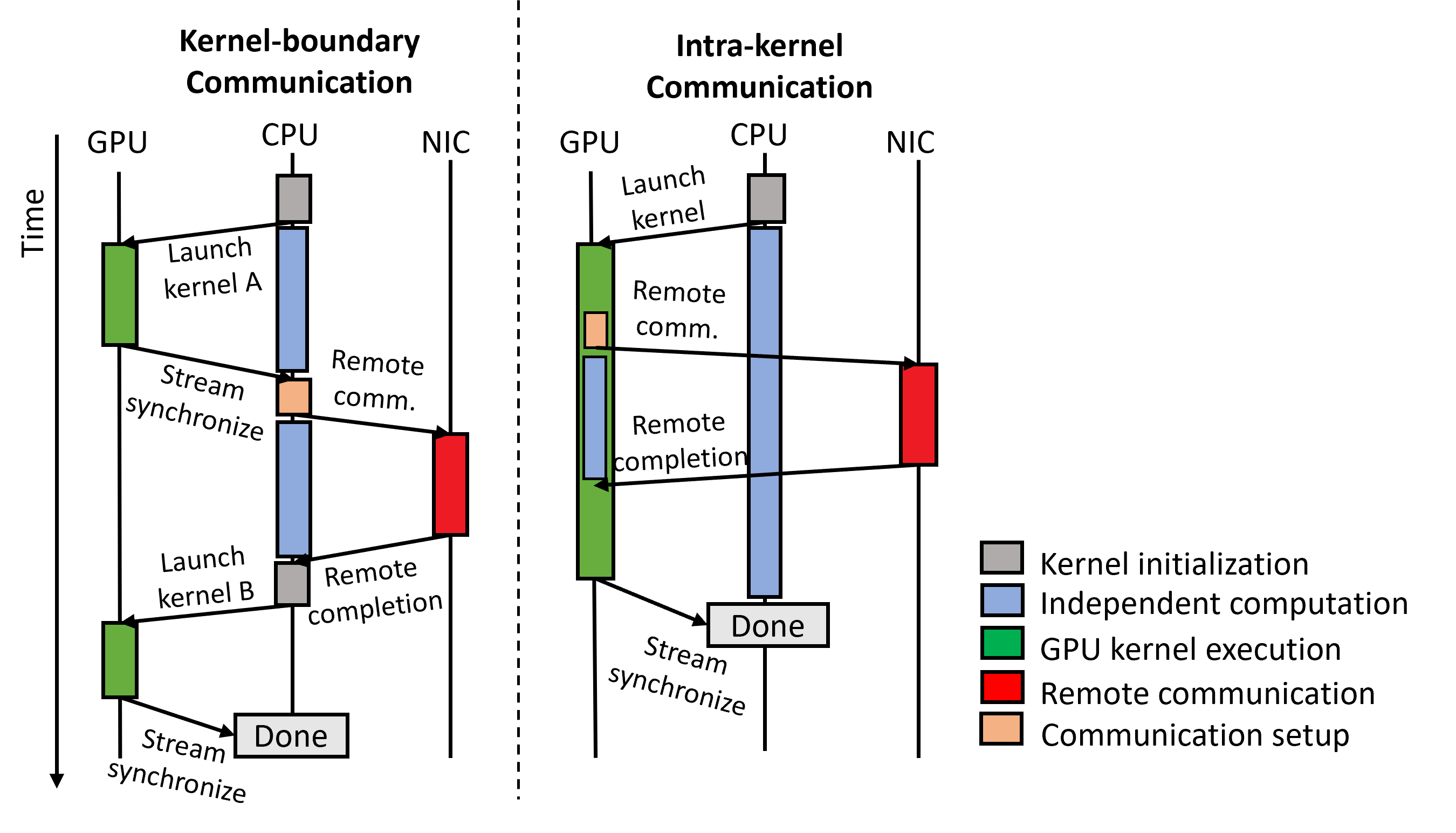}
    \caption{Kernel boundary vs intra-kernel communication}
    \vspace{-10pt}
    \label{fig:gic_transactions}
\end{figure}
\subsection{GPU-initiated intra-kernel communication}
GPU-based HPC and ML systems typically use GPU-Direct Remote Direct Memory Access (RDMA)~\cite{gpudirect} and associated programming models (e.g., CUDA-Aware MPI~\cite{cuda-aware_MPI}) to enable direct data transfers between GPU and NIC. Bypassing CPU memory helps achieve better latency and thus these approaches are widely used in HPC~\cite{cuda-aware_MPI_profile,cuda-aware_MPI_use} and ML (RCCL~\cite{rccl}, NCCL~\cite{nccl}). However, such communication is performed at kernel boundaries typically by the host CPU. 

GPU-initiated intra-kernel communication allows GPU threads to directly initiate communication with NICs and peer GPUs. Vendor-specific GPU libraries (e.g., NVSHMEM~\cite{nvshmem_web}, ROC\_SHMEM~\cite{rocshmem_web}, MSCCL++~\cite{mscclpp}) have been developed that enable applications to perform intra-kernel GPU-initiated communication. 
Moreover, recent GPU micro-architecture features are facilitating GPU-initiated networking further. For example, prior versions of ROC\_SHMEM required data to be allocated as un-cacheable in order to prevent stale data from being communicated to remote nodes. However, the recent introduction of intra-kernel cache flush instructions~\cite{MI200_isa} allows GPU threads to flush data while the kernel executes to initiate network transactions, while allowing data to be cached during computation. Further, new programming abstractions such as threadblock cluster~\cite{hopper_arch} can enable faster coordination between WGs to perform intra-kernel network communication.

Figure~\ref{fig:gic_transactions} compares and contrasts the system interactions during conventional kernel boundary versus intra-kernel communication. Kernel boundary communication requires the entire computation kernel to complete before the remote/network communication can be triggered (usually by the host CPU). On its completion, CPU then launches the dependent kernel.
To overlap communication, techniques such as double-buffering and GPU streams need to be deployed. In the absence of independent work, applications are broken into smaller independent kernels where each smaller kernel is executed as a separate stream and computation of one stream is overlapped with the communication of another. This can result in large number of small kernels and add significant kernel launch and stream management overhead~\cite{meta_dlrm_neo, fleche, mobileGPU_kernelLaunchOverhead}. In contrast, GPU-initiated intra-kernel communication allows GPU threads to issue remote communication while other threads perform their computation, enabling fine-grained communication and computation overlap within a single kernel. Illustration in Figure~\ref{fig:gic_transactions} shows GPUs directly communicating with NICs~\cite{rocshmem_ppopp}, instead GPU threads can alternatively trigger NIC communication using a CPU proxy{~\cite{mscclpp}}.   



\section{Fused Computation and Communication}
In this section, we illustrate our fused communication + computation approaches for both scale-out and scale-up configurations. We discuss the design and implementation of \emph{embedding + All-to-All} fused operator to demonstrate our scale-out approach, while we use \emph{GEMV + AllReduce} operator as an example to explain our scale-up approach which avoids intermediate buffering and copy operations.

\begin{figure*}[t]
\captionsetup[subfloat]{aboveskip=1pt}
\centering
\subfloat[Example embedding + All-to-All operation]{\includegraphics[width = 0.50\textwidth,keepaspectratio]{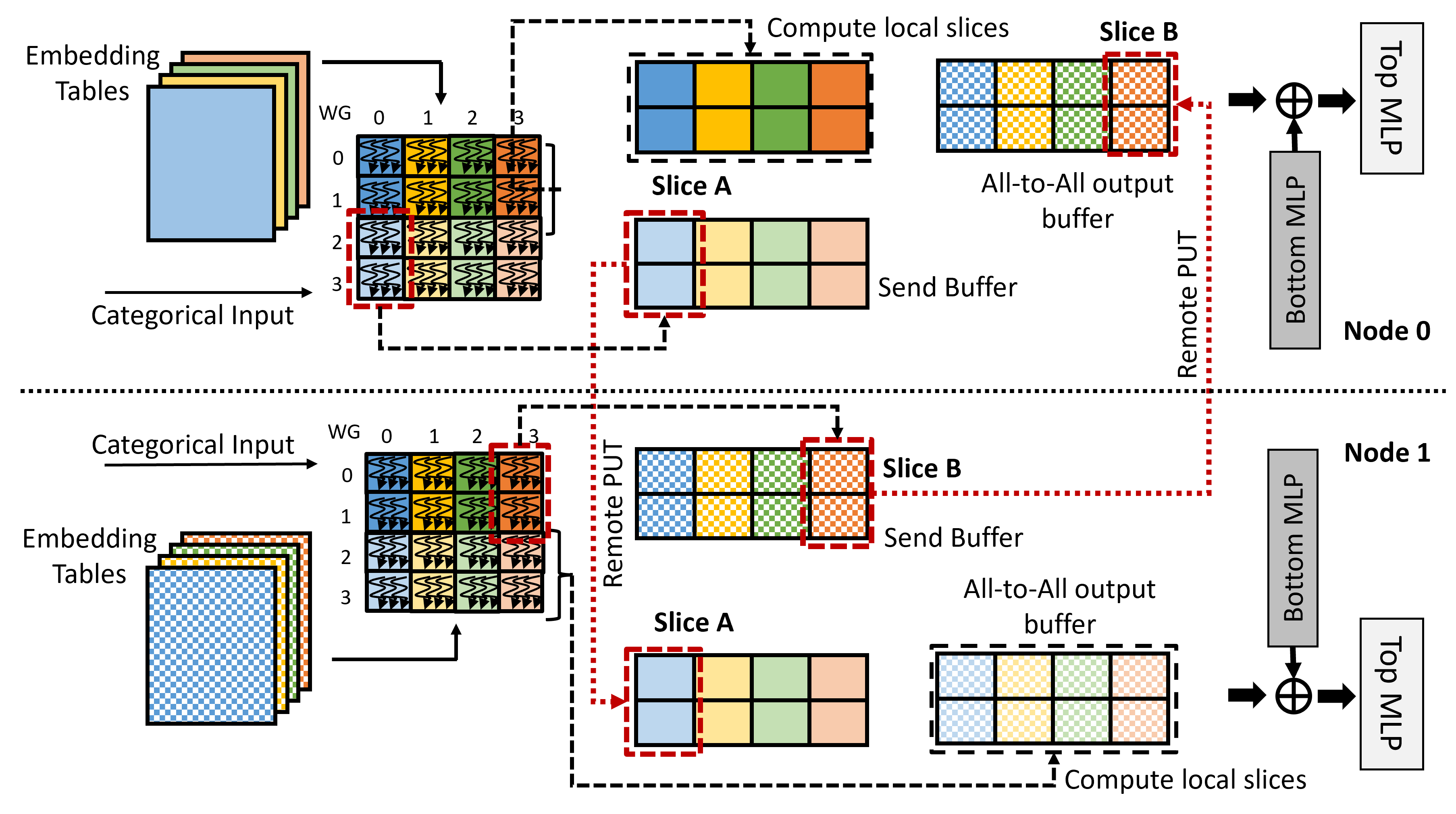} \label{fig:dlrm_fusedOp_example}}
\subfloat[Execution timeline of WGs for Embedding + All-to-All.]{\includegraphics[width = 0.45\textwidth, keepaspectratio]{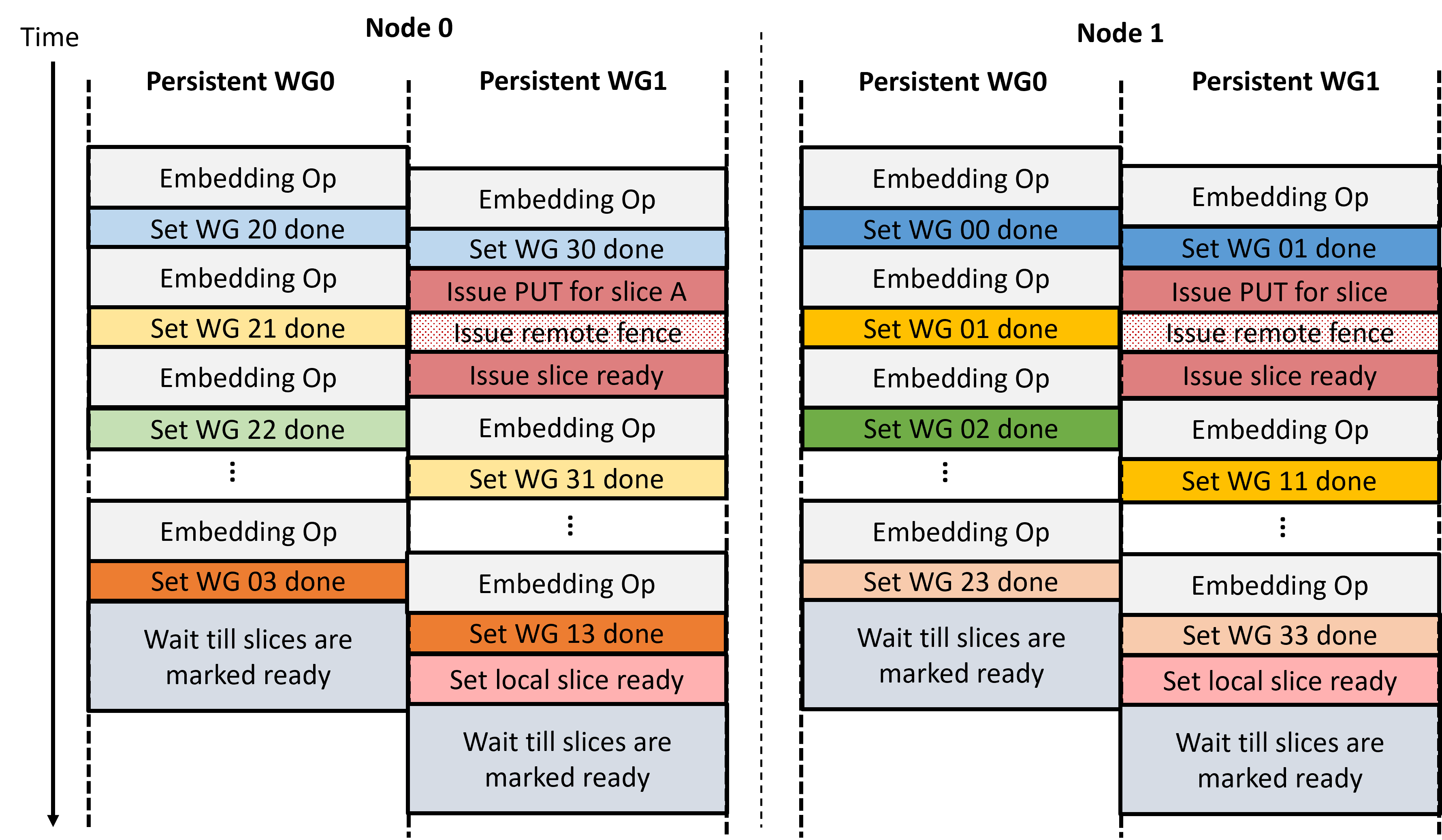} \label{fig:dlrm_timeline}}
\caption{Embedding + All-to-All fused operator.}
\vspace{-10pt}
\label{fig:dlrm_fusedOp}
\end{figure*}

\subsection{Scale-out: Fused Embedding + All-to-All Operator}
We develop the fused embedding + All-to-All operator as a persistent GPU HIP~\cite{HIP} kernel, which performs both embedding pooling (reduction-like) computations and All-to-All communication. We use the ROC\_SHMEM~\cite{rocshmem_web} library to issue intra-kernel communication and 
Figure~\ref{fig:dlrm_fusedOp} shows its execution. It illustrates a two-node system where embedding tables are distributed in model parallel fashion such that there are four tables per node. The All-to-All output and send buffers are allocated within each GPU's symmetric heap (using \emph{roc\_shmem\_malloc()} API). Memory allocated within symmetric heap are registered with the NICs, thus allowing NICs to directly move data between these GPU buffers. We implemented our fused \emph{embedding + All-to-All} operator as a persistent thread kernel~\cite{cta_sched_persistentkernel, persistentkernel_profile} which multiplexes multiple logical embedding pooling operations onto the long-running WGs executing in the GPU. This allows us to schedule logical WGs computing the same slice (output fragment) together (similar to the approach used in ~\cite{cta_sched_persistentkernel}) and further reduces the number of kernel invocations. The kernel is launched with a fixed, input-independent grid size (less than or equal to maximum occupancy as determined from the HIP occupancy API~\cite{hipOccupancyApi}). Each long-running (a.k.a. persistent) WG executes a task loop where every iteration corresponds to the computation performed by a logical WG of the original embedding kernel (\emph{EmbeddingBag\_updateOutputKernel\_sum\_mean}). 

Figure~\ref{fig:dlrm_fusedOp_example} shows the logical WGs and is color coded to indicate which tables are processed by the WGs. Our fused kernel takes in categorical inputs and embedding tables as kernel arguments. The illustrated example assumes a global batch size of four, the slice size to be two embedding output vectors, and that each output embedding vector is computed by one WG. The pooled output embedding vectors from each table are shuffled equally across both nodes, with the first half of the global batch stored in node 0 and the second half in node 1. Depending on their embedding slice, the WGs may need to store their results locally or communicate them to the remote node. The index of the slice computed by a WG can be determined using the output embedding entry and the size of the slice. The slice index along with the global batch size and node count can then be used to determine if the slice needs to be communicated remotely. Logical WGs \emph{WG 00} - \emph{WG 13} compute the output corresponding to the first half of global batch, while \emph{WG 20} - \emph{WG 33} compute the latter half. For node 0, \emph{WG 00} - \emph{WG 13} compute the output entries consumed locally, and results of \emph{WG 20} - \emph{WG 33} will have to be communicated to node 1. While for node 1, results from \emph{WG 00} - \emph{WG 13} are communicated to node 0 and \emph{WG 20} - \emph{WG 33} store their results locally. 

As described earlier, our approach performs communication at a granularity of a \emph{slice}. The size of each slice is set to match the output computed by one or more WGs. In the example shown in Figure~\ref{fig:dlrm_fusedOp_example}, 
\emph{WG 20} and \emph{WG 30} compute slice A on node 0. While there can be multiple WGs responsible for computing a single slice, the remote communication to move the slice is initiated by the last WG that finishes the corresponding slice. In the example shown, assume \emph{WG 30} finished its computation for slice A after \emph{WG 20}. Then, a thread from \emph{WG 30} will issue the remote communication (Remote PUT) to move slice A to node 1. We track the completion of logical WGs corresponding to every slice to identify the last completing WG and trigger communication accordingly. Implementing All-to-All using point-to-point transactions allows the communicating WGs to move the slice data to the destination in a layout required by any subsequent kernel (e.g., the interaction operation in DLRM). The output data can be shuffled at a slice granularity without requiring explicit shuffling or rearrangement. In our approach, we generate output with the shape: \{local batch size, (numTables $\times$ embedding dimension)\}, which can then be passed to the \emph{interaction} operator. 

\emph{Book-keeping Flags:} In our approach, we maintain two sets of flags (not shown in Figure~\ref{fig:dlrm_fusedOp_example}) per GPU to synchronize WG communication and determine the end of communication. First, we maintain a \emph{WG\_Done} bitmask per slice where each bit indicates the completion status of the logical WGs. This bitmask is used to identify the last finishing WG and issue remote communication. Second, we maintain a \emph{sliceRdy} flag per slice (both locally computed and received remotely) to indicate if all the individual slices have been received or computed. The \emph{sliceRdy} flags are set by GPUs computing and communicating the slices and thus are allocated in symmetric memory. The receiver GPUs poll on \emph{sliceRdy} flags to determine if the slices are ready for consumption.

\emph{Communication-aware Scheduling:} Figure~\ref{fig:dlrm_timeline} shows the execution timeline of the persistent WGs for the illustrated example. Consider two persistent WGs per node, where each iteratively performs the work of 16 logical WGs. In our remote communication aware approach, the logical WGs computing slices for remote communication are computed ahead of the WGs computing locally consumed slices, thereby maximizing the opportunity to overlap remote communication. For example, the persistent WGs in node 0 compute \emph{WG 20} and \emph{WG 30} ahead of \emph{WG 00} and \emph{WG 10} so that the remote communication can be overlapped with local slice computation. 

\emph{Synchronization:} Upon completion of each logical WG (an iteration within the task loop), a leader thread sets the bit corresponding to the logical WG in the \emph{WG\_Done} bitmask as shown in the timeline. The WG also checks if it is the last one to complete by testing if all the other bits in the bitmask are set. This design allows the WGs to make forward progress after setting their respective \emph{WG\_Done} flags instead of waiting on an inter-WG barrier. The last completing WG issues two remote PUT calls separated by a remote fence. The first call is to move the slice data, while the second sets its corresponding \emph{sliceRdy} flag on the remote node. The remote fence ensures that the \emph{sliceRdy} flag is only set after the prior PUT has completed. After executing all the logical WGs, the persistent WGs poll on a distinct subset of \emph{sliceRdy} flags before exiting. This ensures that the data from all slices are ready for subsequent kernels while incurring less overhead than having all WGs poll on the entire set of \emph{sliceRdy} flags. Alternatively, synchronization can be performed using a \emph{quiet()} call along with a barrier per GPU.   

\begin{figure*}[t]
\captionsetup[subfloat]{aboveskip=1pt}
\centering
\subfloat[Example GEMV + AllReduce operation.]{\includegraphics[width = 0.3\textwidth,keepaspectratio]{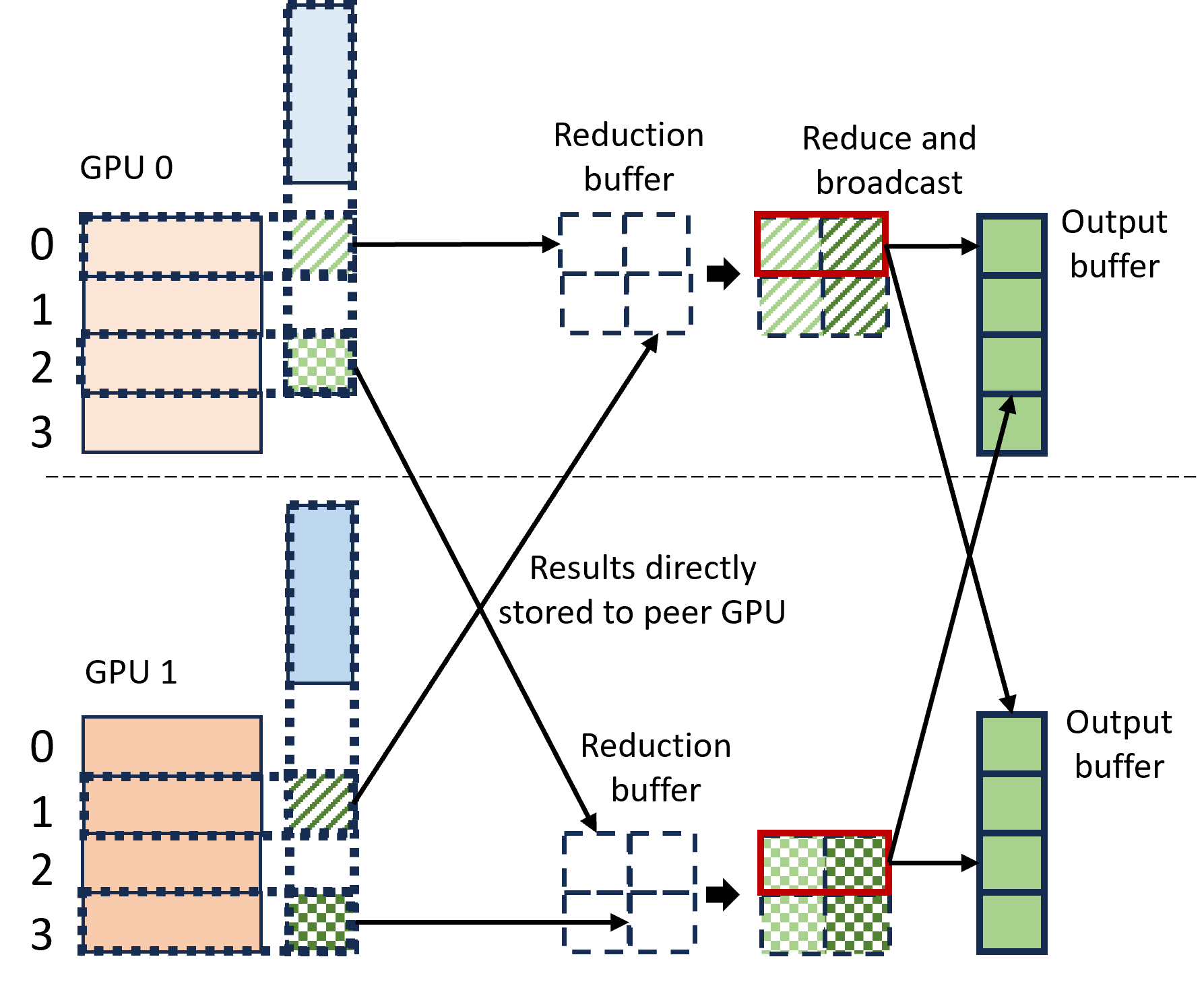} \label{fig:gemv_ar_example}}
\hspace{5pt}
\subfloat[Execution timeline of WGs for GEMV + AllReduce.]{\includegraphics[width = 0.43\textwidth, keepaspectratio]{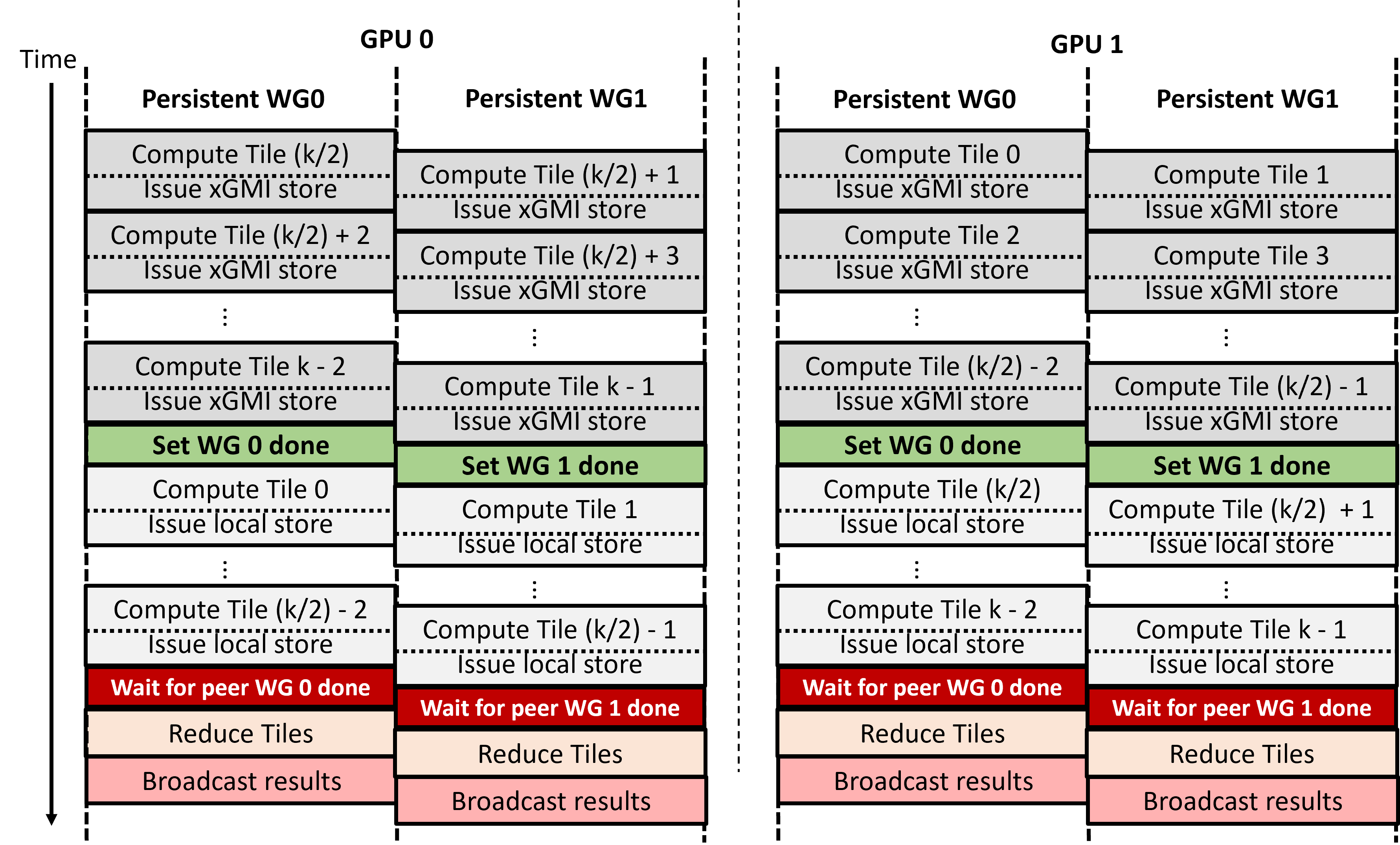} \label{fig:gemv_ar_timeline}}
\caption{GEMV + AllReduce Operator.}
\vspace{-10pt}
\label{fig:gemv_ar_fusedOp}
\end{figure*}

\subsection{Scale-up: GEMV + AllReduce}
Unlike remote GPUs (scale-out) which require RDMA network transactions to communicate, scale-up communication can be performed using native GPU load/store instructions over Infinity Fabric\texttrademark~ connections{~\cite{infinity_fabric}} or NVLink{~\cite{nvlink}}. We exploit this feature to further optimize our fused kernel for scale-up communication. 
We create an optimized zero-copy fused kernel, where in-addition to overlapping dependent communication and computation, we eliminate the stores to intermediate buffers by directly writing the output to the peer GPU memory. The remainder of this section describes the fused \emph{GEMV + AllReduce} operator and contrasts this scale-up approach to our scale-out approach described in the previous subsection. 

Figure{~\ref{fig:gemv_ar_fusedOp}} illustrates our proposed fused \emph{GEMV + AllReduce} operator. It assumes two GPUs connected via Infinity Fabric\texttrademark~ within a single node and each GPU uses a temporary buffer to perform the reduction. We implement our fused operator as a persistent kernel where each physical WG iteratively executes multiple logical WGs responsible for computing individual output tiles. Every physical WG is responsible for computing the same output tiles across GPUs to simplify dependency during reduction. In our approach, we use a two-phase direct algorithm for AllReduce as it has the least number of steps and achieves lower latency for fully connected GPUs {~\cite{NVGroup_allreduce}}. Every GPU is responsible for reducing $\frac{1}{\#GPUs}$ number of tiles (reduce scatter phase) and broadcast the result to other GPUs (all gather phase) as shown in Figure{~\ref{fig:gemv_ar_example}}. In this example, each GPU will compute the entire GEMV output vector but one half of the tiles will be reduced locally while the other will be communicated to peer GPU for the reduction. For example, GPU 0 will reduce the first two tiles, while GPU 1 will reduce the last two. 

We perform communication-aware scheduling of logical WGs similar to scale-out approach, where logical WGs computing tiles which need to be communicated are executed ahead of the logical WGs generating locally consumed tiles. In the example timeline shown in Figure{~\ref{fig:gemv_ar_timeline}}, we can see that WGs in GPU 0 compute and communicate tiles $\frac{k}{2}$ : $k-1$ (k=total output tiles) ahead of tiles 0 : $\frac{k}{2} - 1$ which are locally reduced. The scale-up communication is performed using GPU stores over Infinity Fabric\texttrademark~, so unlike our scale-out approach where a single thread initiated the network communication, all the threads within the WG computing the outputs directly store the results into the destination buffer, eliminating any local copy and intermediate buffering. In order to reduce the amount of synchronization between GPUs, every physical WG only sets one ready flag per peer GPU indicating that the communication of all the tiles (from the specific physical WG) to the peer GPU has finished as shown in Figure{~\ref{fig:gemv_ar_timeline}}. Once the WGs have computed their local tiles, they wait for the flag to be set by the counterpart physical WGs from peer GPUs. On verifying that the flag is set, they reduce the tiles and broadcast the output to other peer GPUs.

Our fused \emph{GEMM + All-to-All} operator also has a similar implementation as explained above except it is implemented in Triton with communication extensions. GPUs perform GEMM operations and the output tiles are communicated to peer GPUs. In this case, no reduction is performed as GEMM operation is fused with All-to-All collective.

\subsection{Overheads}
Having GPU threads initiate communication within kernels consumes GPU resources which can impact performance. This subsection describes those overheads.

\emph{API Latency:} An obvious overhead of having GPU threads initiate networking transactions is the latency of issuing the APIs. The impact of this overhead is limited as the communication is only triggered once per slice. However, there are other book-keeping operations (e.g., setting \emph{WG\_Done} flags) which need to be performed which adds additional overhead. 

\emph{Occupancy:} Invoking ROC\_SHMEM API calls consume GPU registers which otherwise would have been used by the application. Reduced register availability can result in lower occupancy or increased register spilling, impacting performance. Our fused \emph{embedding + All-to-All} kernel achieves 12.5\% lower occupancy compared to the original embedding pooling kernel. Despite lower occupancy, our approach achieves better performance as shown in Section~\ref{sec:eval}.

\emph{Inter-WG Synchronization:} In our approach, communication is triggered once for every slice of data. Since multiple WGs could be computing an individual slice, inter-WG synchronization is required. Our implementation does not use inter-WG barrier instructions, but rather uses cross-lane operations~\cite{crosslane_op} to reduce the \emph{WG\_Done} bitmask and ensure only the last completing WG per slice issues the remote communication. 

\subsection{Integration with ML Frameworks}
While our approach uses the multi-threaded nature of GPU execution to perform fine-grain overlap of collective communication with computation that would not be possible at the kernel granularity, it has to be integrated within existing ML frameworks to be considered a pragmatic solution for wider adoption. In this work, we have implemented two integration approaches and this section describes how these approaches minimize programmers/developers effort. 

\emph{PyTorch Integration:} Our fused \emph{embedding + All-to-All} and \emph{GEMV + AllReduce} operators are implemented in HIP{~\cite{HIP}} and have been integrated within PyTorch. Specifically, we have created new APIs in PyTorch for 1) allocating device memory in symmetric heap and moving a tensor from the CPU's host memory to the allocated device memory (similar to existing $torch.tensor.to()$ API) and 2) launching fused kernel operations (e.g., $torch.embeddingAll2AllOp()$). Such interface reduces programmer overhead as the underlying implementation is hidden from the user. Furthermore, their use can be automated using existing graph transformation optimizations within ML frameworks{~\cite{pytorch_fusion}},{~\cite{tf_fusion}}.

\emph{Extending Triton Framework:}
Effortless development of fused operators is vital for our approach to be widely adopted. We make a crucial step in this direction by extending the Triton{~\cite{triton_paper}} framework to include communication primitives. We chose Triton because it offers a Python-like language which is widely used by ML developers{~\cite{triton}}. In addition, Triton is already integrated with PyTorch{~\cite{triton_pytorch}} allowing code written in Triton to be executed from PyTorch. Triton's Python-like interface enables low-overhead development of new operators. We extended it by creating a Python wrapper for the ROC\_SHMEM library and its scale-up communication APIs and used it to develop our \emph{GEMM + All-to-All} fused operator (our extensions currently support scale-up communication). It is possible to automate the generation of fused kernels using compilers that can track workgroup dependencies across GPU kernels{~\cite{depTrackingCompiler}} and expose communication operations at an intermediate representation level{~\cite{mlir_mpi}}. However, implementing such a compiler flow is beyond the scope of this work.

\section{Evaluation}
\label{sec:eval}
We evaluated the scale-up fused operators on four GPUs connected via Infinity Fabric\texttrademark~, while our scale-out operators are evaluated using both real hardware (two GPU nodes connected via Infiniband) and simulation (128 GPU nodes).

\begin{table}[!t]
\footnotesize
\setlength\extrarowheight{1pt}
\caption{System Setup.}
\label{table:hw_setup}
\centering
\begin{tabular}{||K{1.1cm}|K{6.6cm}||} 
\hline
GPU & AMD Instinct\texttrademark~MI210 \\
\hline
Software & PyTorch v2.0, ROCm v5.4, ROC\_SHMEM v1.6 \\ 
\hline
Scale-up & 4 GPUs (fully connected over Infinity Fabric\texttrademark~(80GB/s)\\
\hline
Scale-out & 2 nodes (with x1 GPU), connected over IB (20 GB/s) \\
\hline
\end{tabular}
\end{table}

\begin{table}[!t]
\footnotesize
\setlength\extrarowheight{1pt}
\caption{Scale-out simulation setup.}
\label{table:astra_setup}
\centering
\begin{tabular}{||K{2.7cm}|K{4.9cm}||} 
\hline
\multicolumn{2}{|| c ||}{\textbf{DLRM Model Parameters~\cite{meta_dlrm_neo}}}\\
\hline
Embedding dimension & 92\\
\hline
MLP layer & Avg. size: 682, \# layers: 43 \\ 
\hline
Avg. pooling size & 70 \\ 
\hline
\multicolumn{2}{|| c ||}{\textbf{ASTRA-Sim Network Parameters~\cite{themis}}}\\
\hline
Topology & 2D Torus (BW: 200Gb/s, lat: 700ns)\\
\hline
\end{tabular}
\vspace{-15pt}
\end{table}

\begin{figure*}[t]
  \centering
     \includegraphics[width =2.1\columnwidth,keepaspectratio]{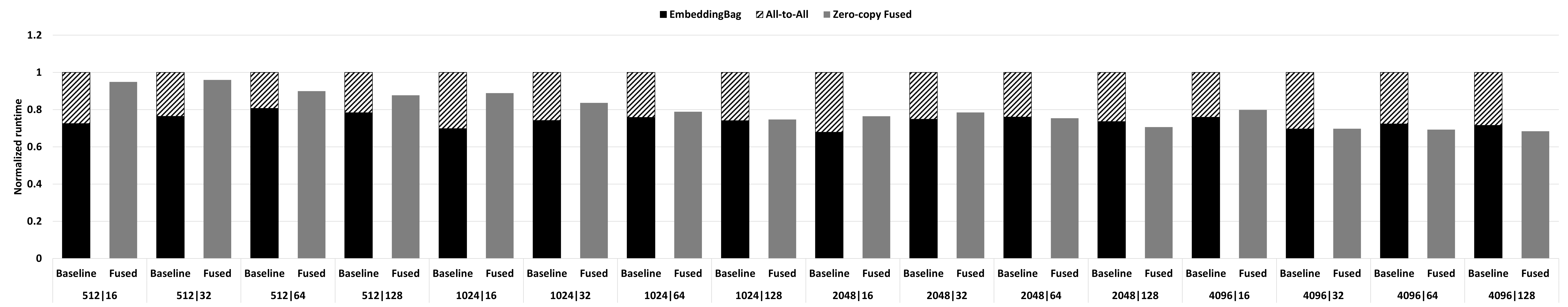}
    \caption{Normalized execution time (intra-node).}
    \vspace{-10pt}
    \label{fig:dlrm_intranode_results}
\end{figure*}
\subsection{Setup}
We compared our fused operators against bulk-synchronous execution of computation and communication kernels (blit kernel-based collectives for intra-node or GPU-Direct RDMA-based collectives for inter-node, both from RCCL{~\cite{rccl}}). Table{~\ref{table:hw_setup}} shows the system setup used for our evaluations.

\emph{Embedding + All-to-All:} We evaluated embedding operations (embedding dim=256{~\cite{meta_dlrm_neo}}) implemented in the public DLRM{~\cite{dlrm_git}} code with All-to-All. In our fused kernel, a cluster of 32 logical WGs compute an output slice to achieve a much higher degree of fine-grain communicaiton-computation overlap than what would be possible by splitting the kernel. For example, batch size=2048, tables/GPU=256 configuration resulting in 512k total logical WGs, will require 16384 additional kernel launches, resulting in such a significant overhead{~\cite{meta_dlrm_neo, fleche, mobileGPU_kernelLaunchOverhead}} that we did not evaluate it.

We varied the global batch-size and the number of embedding tables processed per GPU across different inter-node and intra-node configurations. Each configuration is labeled as: $\{$global batch size $|$ embedding tables per GPU$\}$ in the graphs. For inputs, we used the data generator in DLRM. For large scale-out simulation, we evaluated the entire DLRM \cite{dlrm_git} application in ASTRA-Sim~\cite{astrasim}. We modified ASTRA-Sim's execution graph to model our fused \emph{embedding + All-to-All} kernel. The DLRM model and simulator parameters are shown in Table~\ref{table:astra_setup}. The per-kernel execution times used in ASTRA-Sim were collected on an AMD Instinct\texttrademark~MI210 GPU using ROC-profiler~\cite{rocprof}.

\emph{GEMV + AllReduce:} This operator is specific to inference, so we evaluate them in a scale-up configuration. The computation involved is relatively smaller (inference workload) and thus splitting them into smaller kernels to achieve computation-communication overlap will under-utilize GPUs and exacerbate kernel launch overheads. Thus, we compare our approach against a bulk-synchronous RCCL-based baseline. We select the input matrix and vector dimensions to reflect the sizes in current and future transformer models{~\cite{T3}}.

\emph{GEMM + All-to-All:} The evaluated fused operator is implemented in Triton. We use the publicly available Triton GEMM implementation{~\cite{triton_git, triton_rocm_git}} and modify it to perform All-to-All using our communication extensions. The input matrix sizes are based on commonly seen dimensions in MoE layers{~\cite{tutel}}.

\subsection{Scale-up evaluation}
We evaluated the benefits of the zero-copy, fused kernel on a single node with four GPUs. All communication is performed at the GPU thread granularity (not slice) using stores across the Infinity Fabric\texttrademark~ links.

\begin{figure}[t]
  \centering
     \includegraphics[width =\columnwidth,keepaspectratio]{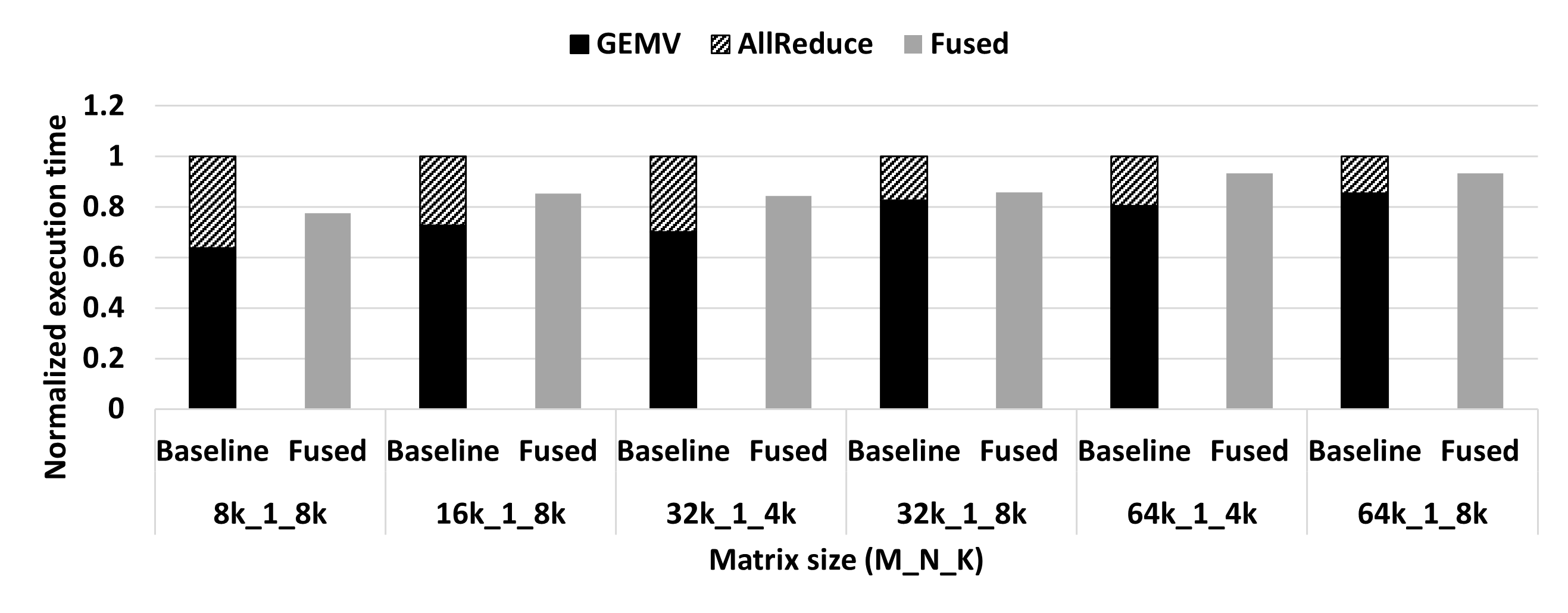}
    \caption{Normalized execution time: GEMV + AllReduce.}
    \vspace{-10pt}
    \label{fig:gemv_ar}
\end{figure}

\emph{Embedding + All-to-All:} Figure{~\ref{fig:dlrm_intranode_results}} shows the execution time of our approach normalized against the time taken by the bulk-synchronous baseline. Overall, our approach achieves on average 20\% (and up to 32\%) lower execution time. The benefits achieved are lower for smaller batch sizes (e.g., 512) due to the small All-to-All latency. While for larger batches, we observe higher benefits than what could be achieved by fully overlapping All-to-All due to the zero-copy optimizations which eliminate stores to intermediate local buffer.

\emph{GEMV + AllReduce:}
Figure{~\ref{fig:gemv_ar}} shows the execution time of our \emph{GEMV + AllReduce} fused operator normalized against time taken for bulk-synchronous execution of GEMV and intra-node AllReduce for different matrix sizes. Our approach achieves on average 13\% (and up to ~22\%) lower execution time. The latency reduction is lower for the larger inputs (M = 64k)  because as the output vector size grows, the contention for the Infinity Fabric\texttrademark~ links increases, lowering the benefits.

\begin{figure}[t]
  \centering
     \includegraphics[width =\columnwidth,keepaspectratio]{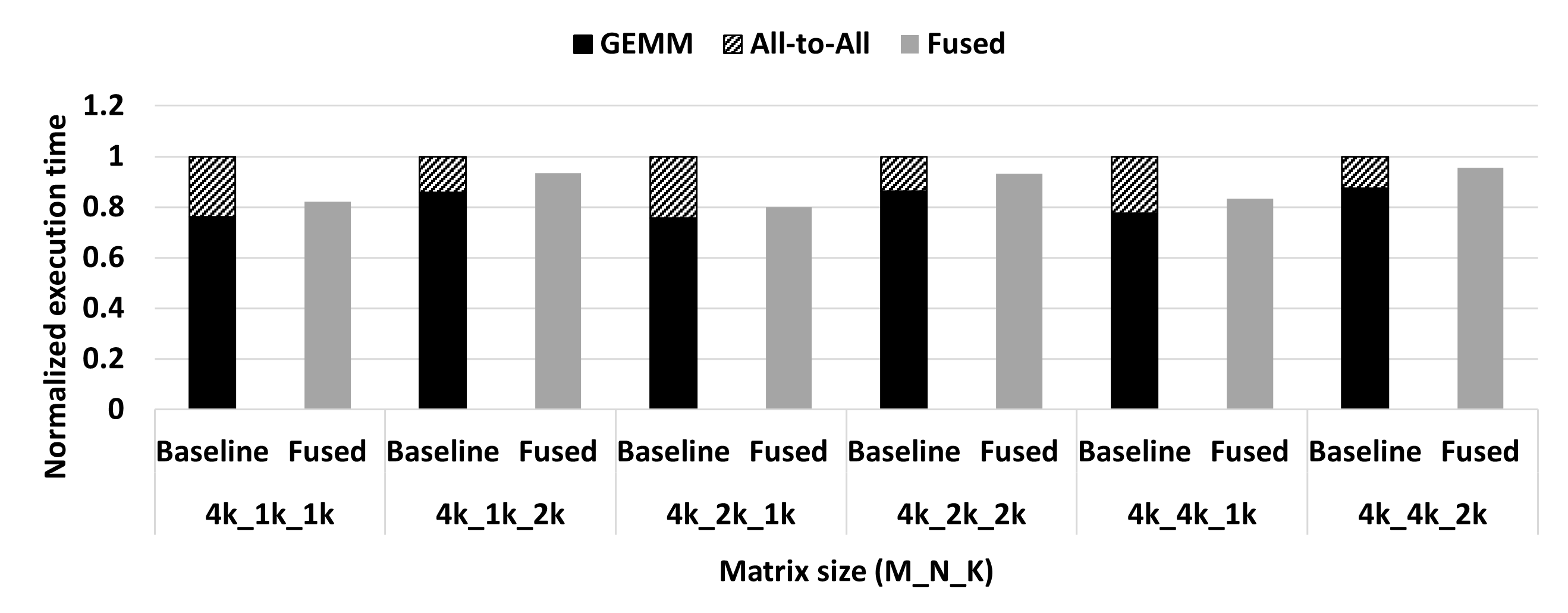}
    \caption{Normalized execution time: GEMM + All-to-All.}
    \vspace{-10pt}
    \label{fig:gemm_a2a}
\end{figure}

\emph{GEMM + All-to-All:}
Figure{~\ref{fig:gemm_a2a}} shows the execution time of our \emph{GEMM + All-to-All} fused operator normalized against time taken for the bulk-synchronous execution of the GEMM and intra-node All-to-All kernels. Our approach lowers execution time by 12\% on average and up to 20\%. Since we are using a generic GEMM implementation provided with Triton, the GEMM dominates the overall execution time and limits the benefits of our approach.  

\subsection{Inter-node evaluation}
\begin{figure}[b]
\captionsetup[subfloat]{aboveskip=-10pt}
\vspace{-10pt}
\centering
\subfloat{\includegraphics[width = 0.35\textwidth,keepaspectratio]{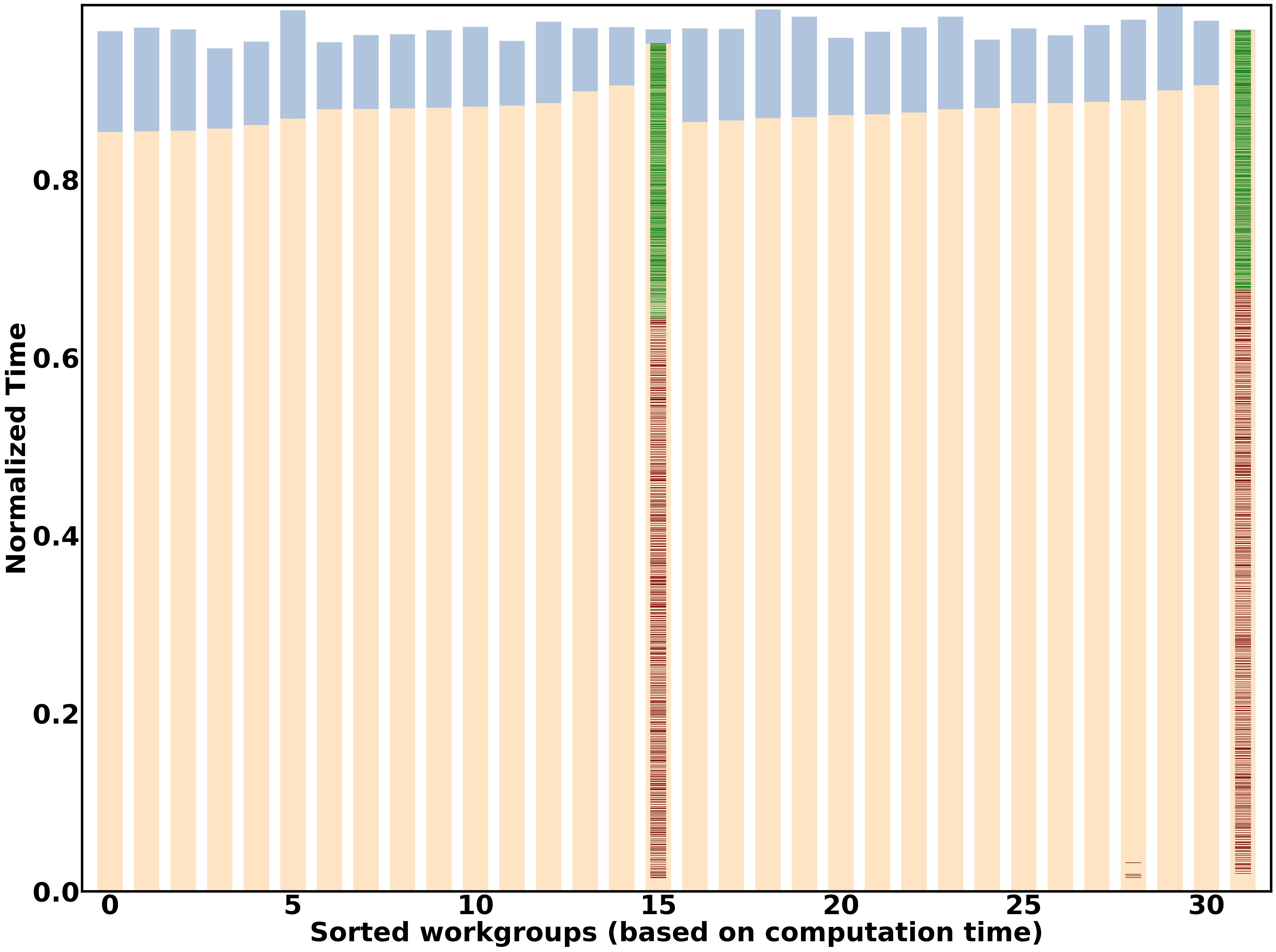}}
\hspace{2pt}
\subfloat{\includegraphics[width = 0.2\columnwidth]{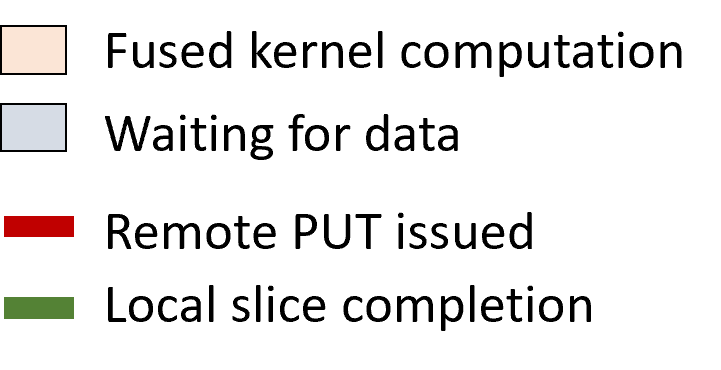}}\\
\caption{Profiled timeline of persistent WG.}
\label{fig:wg_profiled}
\end{figure}

\emph{WG Profiling:}
We first demonstrate the effectiveness of overlapping embedding operations with All-to-All communication. We profile the persistent WGs' execution timeline for a configuration with batch size = 2048, tables/GPU = 256, and with slice size such that each slice is computed by 16 WGs. WGs computing the same slice (cluster of 16) are sorted based on their time to complete embedding pooling operations (earliest first). Figure~\ref{fig:wg_profiled} illustrates the execution timelines for the first 32 persistent WGs along with the points in time when non-blocking network transactions were issued. Figure~\ref{fig:wg_profiled} also shows the instances when the locally consumed slices are computed (marked as \emph{Local slice completion}) to illustrate the effect of communication-aware scheduling. We can see that some persistent WGs are issuing remote communication while others are performing the computation, demonstrating our approach achieves fine-grained communication-computation overlap. Furthermore,  asynchronous network transactions allow the communicating WGs (e.g., WG 15 and 31) to continue performing embedding pooling without being blocked. Figure~\ref{fig:wg_profiled} shows that the last WGs (WG 15 and WG 31) within each cluster of 16 WGs issues most of the communication. This is because each persistent WG iteratively computes individual slices. The last completing WG for one slice often stays the last across multiple slices (recall that only the last completing WG issues the remote non-blocking PUT). Additionally, we can see that the remote communication calls are issued before computing the local slices, maximizing the opportunity to hide remote communication. The time spent waiting on data differs across WGs because each WG waits for a distinct subset of the \emph{sliceRdy} flags.    

\begin{figure*}[t]
  \centering
     \includegraphics[width =2.1\columnwidth,keepaspectratio]{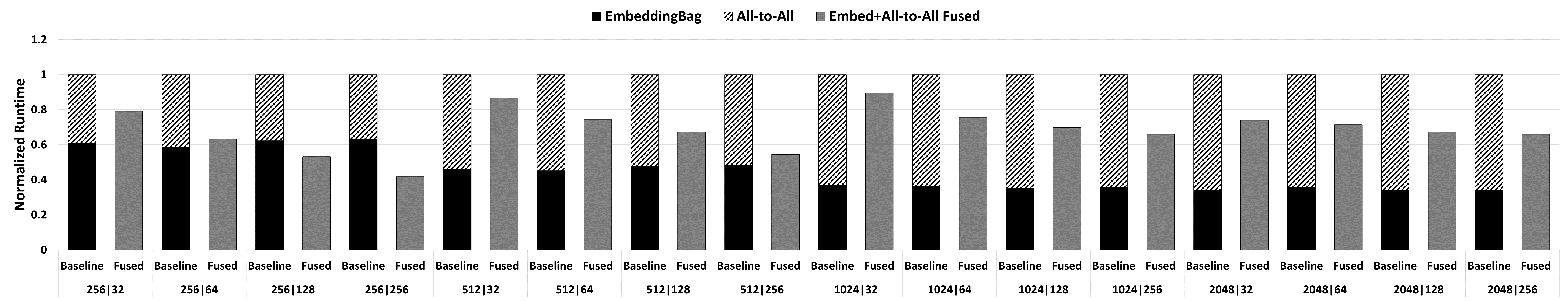}
    \caption{Normalized execution time (inter-node).}
    \vspace{-10pt}
    \label{fig:embedbag_all2all_results}
\end{figure*}
\emph{Execution time:}
Figure~\ref{fig:embedbag_all2all_results} shows the time taken by our fused embedding + All-to-All kernel normalized to the baseline executing separate embedding operation and All-to-All collective kernels. Our fused kernel uses a slice size of 32 embeddings and Figure~\ref{fig:embedbag_all2all_results} demonstrates that our approach benefits a wide set of batch sizes and embedding table counts, achieving 31\% average reduction in execution time (up to 58\%). For smaller global batch sizes (e.g., 256), we note that the performance benefit is more than what could be achieved by fully overlapping communication. This is because smaller batch sizes result in poor compute utilization for the baseline, while our approach achieves much higher compute utilization by processing all tables within a single fused kernel.

\begin{figure}[b]
  \vspace{-20pt}
  \centering
     \includegraphics[width =0.7\columnwidth,keepaspectratio]{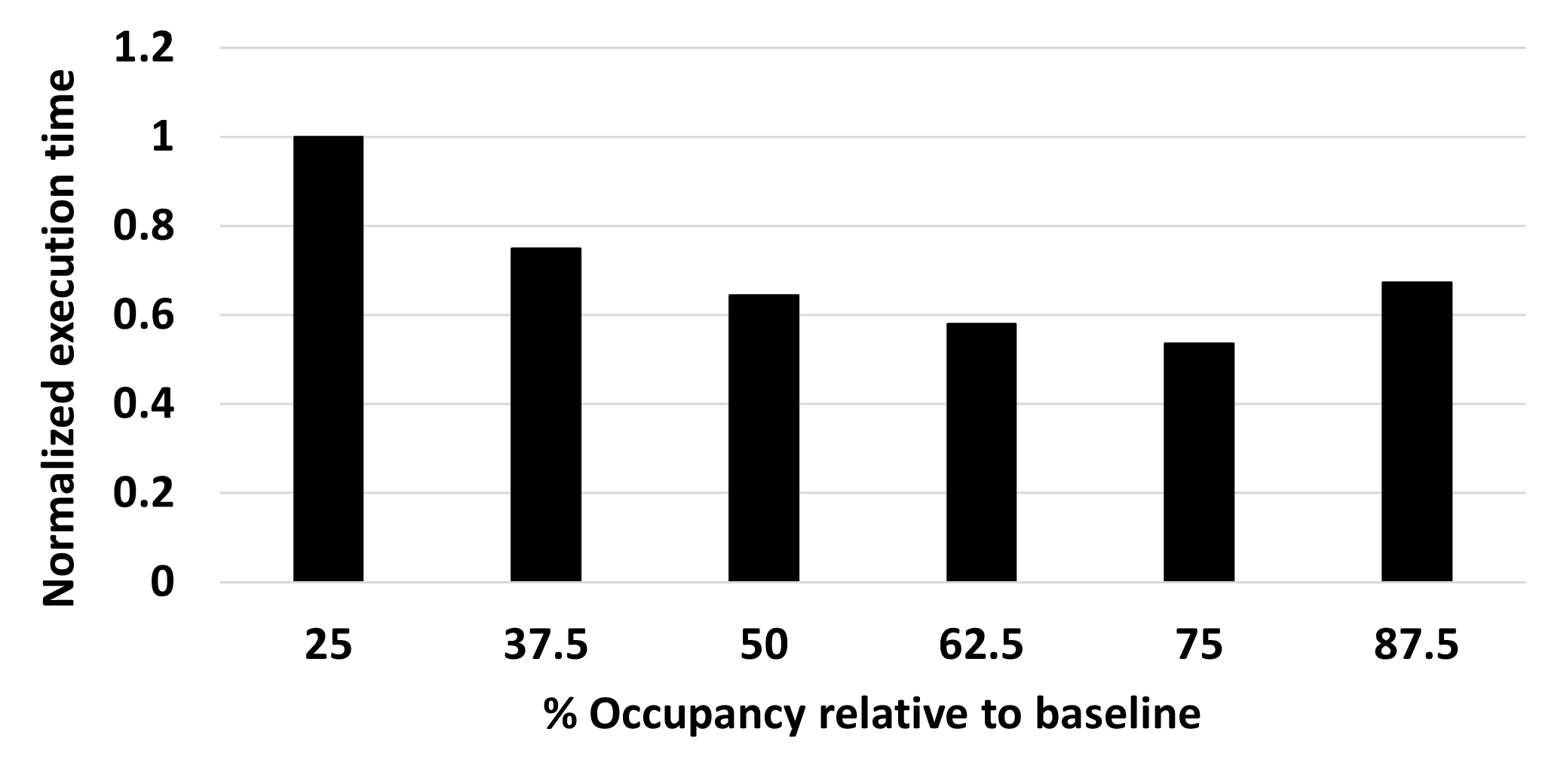} 
    \caption{Impact of WG occupancy on execution time.}
    \label{fig:occ_sweep}
\end{figure}
\emph{Occupancy effect:}
Our fused kernel implementation results in 12.5\% lower GPU occupancy than the baseline due to the extra registers required by GPU-initiated networking operations. However, this loss of occupancy does not degrade performance. Figure~\ref{fig:occ_sweep} shows the variation in the execution time  (global batch size: 1024, tables/GPU: 256) across different GPU occupancy. The figure only shows results up to 87.5\% occupancy because that was the maximum achievable occupancy relative to baseline. We see that as the occupancy is increased from 25\% to 75\%, the parallelism increases and consequently the execution time reduces by 46\%. However, increasing the occupancy further from 75\% to 87.5\% results in the execution time increasing by 25\%. At this higher occupancy level, the memory intensive embedding operations encounter significant memory contention, exemplifying the trade-off between parallelism and memory contention. 

\begin{figure}[b]
  \vspace{-20pt}
  \centering
     \includegraphics[width =\columnwidth,keepaspectratio]{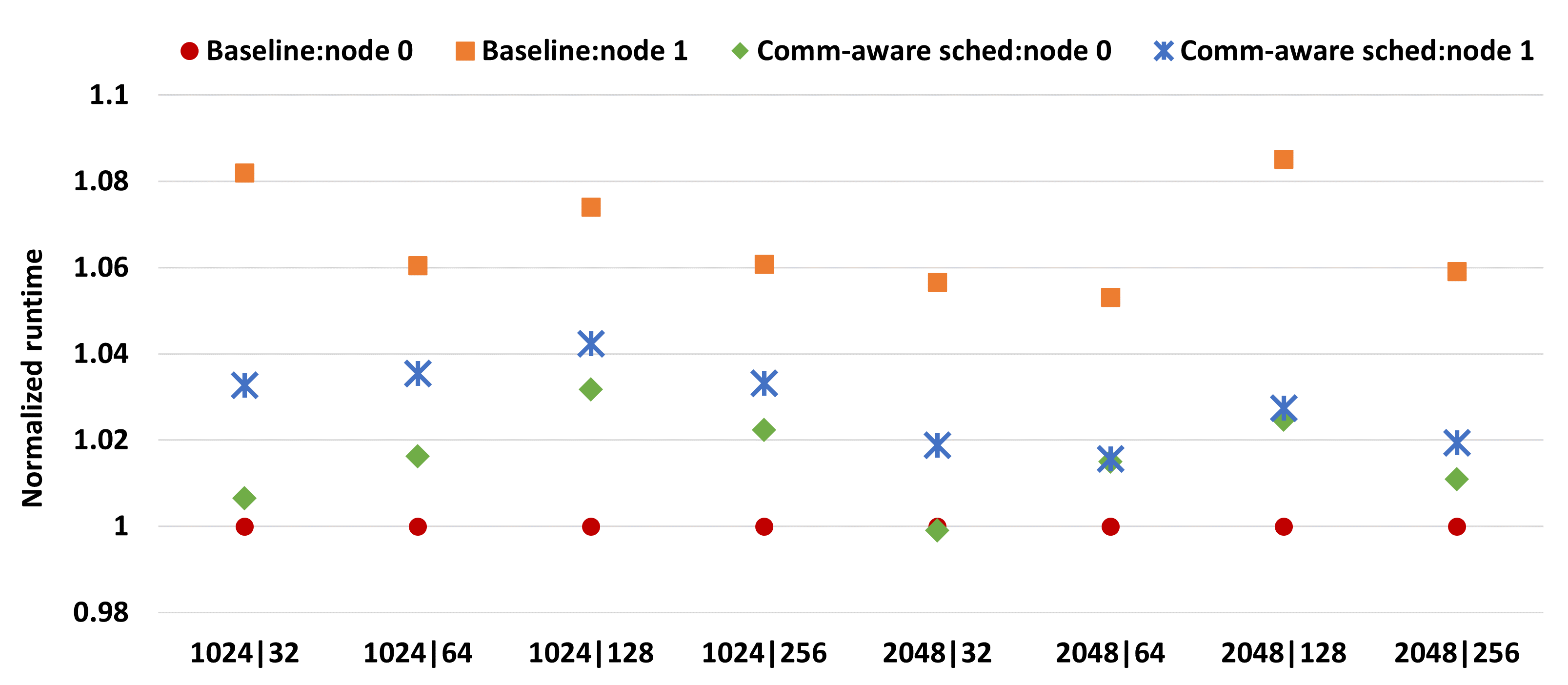}
    \caption{Impact of communication-aware WG scheduling.}
    \label{fig:comm-aware_sched}
\end{figure}

\emph{Communication-aware Scheduling:}
In our approach, we implement communication-aware WG scheduling to maximize the opportunity to hide remote communication. Figure~\ref{fig:comm-aware_sched} shows the fused \emph{embedding + All-to-All} kernel execution time of both nodes (normalized to baseline:node 0) with and without communication-aware scheduling. The baseline communication-oblivious scheduling starts from WG (0,0,0) and then proceeds sequentially. In baseline scheduling, node 0 and 1 have an average execution time skew of $\sim$7\% while using communication-aware scheduling exhibits only a $\sim$1\% average execution time skew. Many distributed ML models, including DLRM, periodically synchronize across nodes (e.g., during synchronous gradient descent), and thus execution skew can reduce the overall performance. The higher skew in the baseline is due to the higher execution time for node 1. As part of our fused kernel implementation, WGs in node 1 will wait for slices from node 0 and vice-versa. Node 1 has the same logical WG schedule under both strategies, where it first computes the slices to be communicated to node 0 (please see Figure~\ref{fig:dlrm_timeline}). However, with communication-oblivious scheduling, node 0 computes the slices to be remotely communicated after the locally consumed slices. Thus the remote communication is not hidden behind local slice computation, delaying the completion of node 1.

\subsection{Large scale-out evaluation}
\begin{figure*}[t]
  \centering
     \includegraphics[width =2\columnwidth,keepaspectratio]{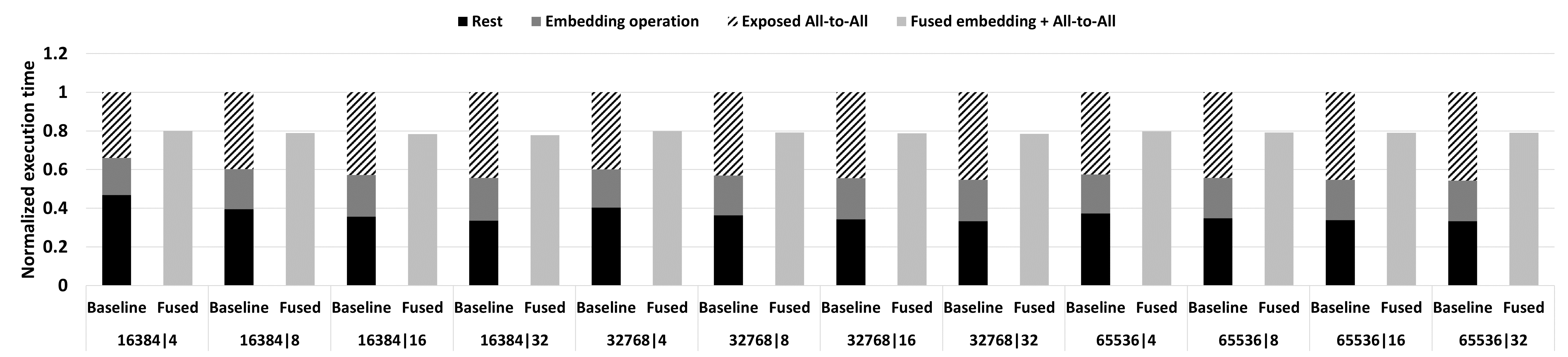}
    \caption{Scale-out simulation normalized execution time.}
    \vspace{-10pt}
    \label{fig:scale_out_results}
\end{figure*}

We evaluated the benefits of our approach on DLRM training distributed across 128 nodes (with one GPU each) using ASTRA-Sim. Figure~\ref{fig:scale_out_results} shows the normalized execution time for performing one DLRM training pass. The embedding operations in both the forward and backward passes are overlapped with their dependent All-to-All collective operations using WG-level parallelism. We can see that our approach is able to hide most of embedding operations , achieving on average a $\sim$21\% reduction in the overall execution time.        

\section{Related Work}
In this section, we discuss prior research done on optimizing distributed ML models, overlapping computation with CPU-initiated inter-node and DMA-assisted intra-node communication. We also discuss previous work on GPU-initiated communication.    

T3{~\cite{T3}} proposes hardware track and trigger mechanisms to detect GPU stores and initiate intra-node DMA communication to overlap it with computation. While this work aims to achieve communication-computation overlap, it requires hardware modifications to track GPU stores using a new widget invisible to programmers that can be difficult to debug. In contrast, our approach requires no hardware changes and has been implemented in commercially available frameworks.    

Wang et. al~\cite{google_shard_overlap} propose decomposing the original collective communication and its dependent computation into smaller pieces. The communication of an individual data shard can be overlapped with the computation of another. Our approach is conceptually similar to this as both aim at overlapping communication of a portion of data with the computation of another. However, unlike us, Wang et al.'s approach will result in a higher number of kernel invocations where each sharded kernel is smaller than the original one. For their use case ~\cite{google_shard_overlap}, the sharded kernels are sufficiently large to amortize the kernel invocation overhead but this is not always the case~\cite{meta_dlrm_neo}. Further, their approach requires additional operations to combine the individual partial results. Jangda et. al~\cite{coconet} also proposes similar optimizations but they use a new domain-specific language to express ML programs in the form of computation and communication operations. Our approach instead uses GPU-initiated networking integrated within existing ML frameworks, to hide communication without requiring additional kernel launches, and computations.

Mudigere et al.~\cite{meta_dlrm_neo} proposed Neo system to improve distributed DLRM execution. This system employs 4D parallelism strategy (table-wise, row-wise, column-wise, and data parallelism) for distributing embedding computations evenly across GPUs. Furthermore, they fuse compute kernels to minimize kernel-launch overheads and memory requirements, and employ software-managed caching to leverage hardware memory hierarchy. However, in this paper, we try to address the large All-to-All communication latency exposed in distributed DLRM training by fusing and overlapping All-to-All with embedding operations.

Wang et. al.~\cite{gnn_nvshmem} propose using GPU-initiated networking to hide remote access latency in GNNs. While this work uses GPU-initiated communication, it focuses on optimizing irregular communication by interleaving local neighbor aggregation with remote neighbour access. Their work targets a different problem than that addressed in this paper and requires a completely different implementation. 

Unlike our approach which aims at hiding the collective communication latency, there have been efforts to reduce them. Cai et. al.~\cite{msccl} have proposed an approach to synthesize collective algorithms tailored for a specific hardware topology. ARK{~\cite{ark}} proposes offloading intra-node communication to DMA driven by GPU to reduce control-plane overhead and data-plane interference. Others~\cite{themis, multitree_allreduce} have proposed optimal communication schedule for specific collectives aimed at reducing contention and improving link utilization. These optimizations are orthogonal to our approach.   


\section{Conclusion}

In this paper, we hide the collective communication with dependent computation at a fine granularity. GPU-initiated intra-kernel communication
enables wavefront/warp-, WG- or WG-cluster granular communication unlike CPU-initiated networking which provides only kernel-granular communication. 
We developed prototype fused \emph{embedding + All-to-All}, \emph{GEMV + AllReduce} and \emph{GEMM + All-to-All} kernels where fragments of computed outputs are communicated in parallel to other WGs performing the remainder of the computation. We propose remote communication-aware scheduling where the logical WGs computing the remote slices are executed before the WGs computing the local slices, maximizing the opportunity to overlap remote communication. We further optimize scale-up communication among GPUs by developing zero-copy fused kernels. Here, GPU threads perform the computation and store the results directly to the destination address at the peer GPU, thereby eliminating intermediate stores to local memory. We expose our fused operators as new PyTorch operators as well as develop fused operators (\emph{GEMM + All-to-All}) using our Triton framework extensions which include communication primitives to demonstrate integration with ML frameworks.

We evaluated our approach both on actual hardware and using simulation. Our scale-up evaluations show that zero-copy fused \emph{embedding + All-to-All} kernel on average achieves 20\% (and up to 32\%) lower execution time than the baseline. Our \emph{GEMM + All-to-All} and \emph{GEMV + AllReduce} evaluations show that our approach achieves 13\% (up to 22\%) and 12\% (up to 20\%) lower execution time  respectively. 
Our inter-node evaluations show that the execution time taken by fused kernel \emph{embedding + All-to-All} is on average 31\% (and up to 58\%) lower than that of baseline embedding operations and All-to-All collective.
We evaluated the impact of proposed communication-aware WG scheduling and show that it achieves 6\% lower execution skew than communication-oblivious scheduling. Finally, we used ASTRA-Sim to perform large scale-out simulations to evaluate the benefits of our approach on the entire DLRM training run. Our evaluations show that using fused embedding + All-to-All kernels reduce the training time by ~21\% for a 128 node system.

\bibliographystyle{plain}
\bibliography{references}

\end{document}